\documentclass[12pt]{article}
\usepackage{amsmath,amsthm,amssymb,euscript,epsf,epsfig}
\usepackage{array}
\usepackage{fancybox}

\newcommand{\be}{\begin{equation}}
\newcommand{\ee}{\end{equation}}
\newcommand{\bea}{\begin{eqnarray}\displaystyle}
\newcommand{\eea}{\end{eqnarray}}
\newcommand{\nnm}{\nonumber}

\makeatletter
\@addtoreset{equation}{section}
\makeatother

\def\one{{\hbox{ 1\kern-.8mm l}}}
\def\zero{{\hbox{ 0\kern-1.5mm 0}}}

  \def\cO{{\cal O}}

\def\vmn{ V^{\otimes m } \otimes \bar V^{\otimes n } }

\begin{document}
{}~
{}~
\hbox{QMUL-PH-07-19}
\break

\vskip .6cm

\centerline{{\LARGE \bf  Branes,  Anti-Branes
 and Brauer Algebras  }} 
\centerline{{\LARGE \bf in  Gauge-Gravity duality }}

\medskip

\vspace*{4.0ex}

\centerline{ {\large \bf Yusuke Kimura}\footnote{y.kimura@qmul.ac.uk}
{ \bf  and }  {\large \bf Sanjaye Ramgoolam}\footnote{s.ramgoolam@qmul.ac.uk}  } 
\vspace*{4.0ex}
\begin{center}
{\large Department of Physics\\
Queen Mary, University of London\\
Mile End Road\\
London E1 4NS UK\\
}
\end{center}

\vspace*{5.0ex}

\centerline{\bf Abstract} \bigskip

 We propose gauge theory operators built using a complex Matrix scalar
 which are dual to brane-anti-brane systems 
 in $AdS_5 \times S^5 $, in the zero coupling 
 limit of the  dual Yang-Mills. 
 The branes involved are half-BPS giant gravitons. 
 The proposed operators dual to  giant-anti-giant configurations 
 satisfy the appropriate orthogonality properties. Projection operators 
 in Brauer algebras are used to construct the relevant multi-trace 
 Matrix operators. These  are related to  the 
``coupled representations'' which appear in 2D Yang-Mills theory. 
 We discuss the implications of  these results for the quantum
 mechanics of a complex matrix model,  the counting of non-supersymmetric
 operators and the physics of brane-anti-brane systems. 
 The stringy exclusion principle known from the properties of 
 half-BPS giant gravitons, has a  new incarnation in this context. 
 It involves a qualitative 
 change in the map between  brane-anti-brane states to
 gauge theory operators. In the case of a pair  of sphere giant 
 and anti-giant this change occurs when the sum of the 
magnitudes of their angular
 momenta reaches $N$.

\thispagestyle{empty}
\vfill

\eject


\section{Introduction}

 The two-point functions of  gauge
 theory operators in $N=4$ $U(N)$ super-Yang-Mills gauge theory 
 corresponding to highest weights of 
 half-BPS representations can be diagonalised \cite{cjr}.  
 The elements of the diagonal basis
 are given in terms of $ \chi_R ( \Phi  ) $ 
 where $R$ is a Young Diagram of $n$ boxes 
\bea\label{schurop}  
\chi_R ( \Phi ) =  { 1 \over n ! } 
\sum_{ \sigma \in S_n } \chi_R ( \sigma ) tr ( \sigma \Phi ) 
\eea
$ \chi_R ( \sigma ) $ is the character of $ \sigma $ in the 
representation of $S_n$ labelled by $R$.  
$ \Phi $ is a complex matrix which can be viewed as an operator 
acting on an $N$-dimensional vector space $ V $, i.e 
$ \Phi : V \rightarrow V $. It can be extended to give an 
operator transforming $ V^{\otimes n } \rightarrow  V^{\otimes n } $ 
by considering $ \Phi \otimes \Phi  \cdots \otimes \Phi $. 
In the RHS of (\ref{schurop}) the trace is being taken in 
$ V^{\otimes n } $, and  
$ \sigma $ acts on $ V^{\otimes n } $ 
by permuting the factors. From these facts it follows that 
\bea 
tr ( \sigma \Phi ) = \Phi^{i_1 }_{i_{\sigma(1) } } \cdots
 \Phi^{i_n }_{i_{\sigma(n) } }
\eea 
The operator $ \chi_R ( \Phi ) $ can also be viewed as a 
holomorphic continuation of the $U(N)$ character $ \chi_R ( U  ) $
by replacing the unitary matrix $U$ with a complex matrix $ \Phi $. 
It is also useful to view it as a trace  $tr_n ( p_R \Phi ) $
in $V^{\otimes n } $ obtained by using  a projection operator $p_R$
in the group algebra of $S_n $ 
\bea 
p_R = { d_R \over n ! }  \sum_{\sigma \in S_n } \chi_R ( \sigma ) \sigma 
\eea

The 2-point function in this basis of operators  is diagonal 
\bea 
< \chi_R ( \Phi^{\dagger}(x_1)  ) \chi_S ( \Phi(x_2) )  > = 
{ \delta_{R S } f_R \over ( x_1 - x_2 )^{2n} } 
\eea 
where $f_R $ is a simple group theoretic quantity. 
This is derived using the basic formula 
\bea\label{bas2pt}  
 <   \Phi^{\dagger  ~ i }_j (x_1)  \Phi^k_l  (x_2) > = 
 { \delta^{i}_l \delta^{j}_k \over (  x_1 - x_2 )^2 } 
\eea 
In most of this paper we will not be interested 
in the position dependences, so  we will 
drop the $x$'s.
Having a diagonal basis in the space of half-BPS operators allows 
an identification of gauge theory operators corresponding to
half-BPS giant gravitons  \cite{mst,gmt,hhi} 
in $ AdS_5 \times S^5 $
 space-time via the AdS/CFT duality \cite{malda,gkp,witten}.  
Some further aspects and developments related to 
half-BPS giant gravitons  are in 
 \cite{cjr,bbns, cr,ber,silva,llm,taktsu,dms,yoneya,bglm}
and references therein.

  For an appropriate choice of $R$,  $ \chi_R ( \Phi ) $ 
  is dual to a sphere-giant graviton, which is a 
 spherical three-brane moving in $ S^5$. As we will explain, by replacing 
 the 3-brane with an anti-3-brane, and at the same time 
 reversing the direction of rotation, we also have 
 a solution of the same energy. This anti-giant is 
 dual to  $ \chi_R ( \Phi^{\dagger}  ) $. The same remark applies
 to AdS-giants (also known as dual giants). 
 The main 
 interest in this paper is to investigate candidates 
 for systems of giant and anti-giants. This requires 
 a diagonalisation of the two-point function in the space of operators 
 built from both $ \Phi $ and $ \Phi^{ \dagger}  $. 
 This problem can be solved elegantly in terms of  
 {\bf Brauer algebras}. These algebras are parametrised by 
 two positive integers. For the 
 case of $ m $ copies of $ \Phi $ and $n$ copies of $ \Phi^{\dagger} $
 the relevant algebra is $ B_{N} ( m , n ) $. The  associative algebra 
 $ B_{N} ( m , n ) $ contains the group algebra of the 
 product of symmetric groups  $ S_m \times S_n$,
 which is denoted as 
$ \mathbb C [ S_m \times S_n ]$.

 An outline of the proposal for gauge-theory duals of giant-anti-giants 
 and the role of Brauer algebras will be given in section 2. 
 These Brauer algebras will be introduced  more systematically, 
 their   relevance and useful properties explained 
 in section 3. Of particular interest in constructing duals 
 of brane-anti-brane composites will be a subset of  the 
 {\bf orthogonal central projectors} 
 in the Brauer algebra. By central, we mean that 
 the projectors commute with the Brauer algebra.
 Section 4 will be devoted to techniques for the explicit 
 construction of these orthogonal projectors. 
 Examples of these projectors will be given in section 5.

 The gauge invariant operators constructed from 
 central Brauer projectors do not exhaust 
 the complete set of gauge invariant operators
 that can be constructed from $ \Phi $ and $ \Phi^{\dagger}$. 
 To get the complete set we need to consider 
{\bf symmetric Brauer elements}.
The counting of the gauge invariant operators 
in the limit where the Matrices $ \Phi $ is large is 
 known to be given by  Polya theory. 
 In section \ref{sec:countingoperator},  we relate 
the Polya counting to Brauer algebras. 
In section \ref{sec:orthogonalsetoperator}
 we describe an orthogonal basis in the space of symmetric 
 Brauer elements and show how they lead to a diagonal 
 basis for the two-point functions of  multi-trace operators.  
 A  physical interpretation in terms of brane-anti-branes 
 of the orthogonal basis of multi-trace operators is discussed 
 in section 8.  This includes a discussion of an interesting finite 
 $N$ effect we describe as the 
{\bf nonchiral stringy exclusion principle}.

The reader is not assumed to have any prior knowledge
 about Brauer algebras. A summary of useful results 
is given in section 3, along with references to the mathematical 
literature.  For a reader with interest in Brauer algebras, 
 we point out the new formula for dual Brauer elements (\ref{bstar1}).  
The explicit formulae for projectors in section 4  and 5
and the connection with Polya theory of sections 3.4 and 
\ref{sec:countingoperator} 
should also be of interest from this mathematical point of view. 
For the reader familiar with the large $N$ expansion 
of two-dimensional Yang-Mills 
we would point to the new formula for coupled dimensions
 (\ref{newcoupdimform}) as an appetiser.  Explicit examples of 
 orthogonal bases of multi-matrix  operators are given in the appendices.

\section{ Proposal for gauge theory duals of 
brane-anti-brane systems  : Outline } 

 Giant 3-brane gravitons are dual to $ \chi_R ( \Phi ) $. 
   A simple inspection of the derivation of the giant graviton solutions 
 of \cite{mst} shows that   spherical  anti-3-branes
 can provide supersymmetric solutions 
  with opposite angular momentum. When we change the angular velocity 
  $ \dot \phi $ to 
  $ - \dot \phi $ while changing the sign of the Chern-Simons coupling to 
 the background flux, as appropriate for changing brane to anti-brane, 
 the effective Lagrangian and Hamiltonian are unchanged, while the
 angular momentum changes sign.  This leads to the 
 conclusion that anti-branes also provide supersymmetric solutions.
 Giant anti-3-brane gravitons are dual to $ \chi_S ( \Phi^{\dagger} ) $. 
 The  brane-anti-brane composites will  be non-supersymmetric.

 Gauge theory operators dual to brane-anti-brane systems 
 involve both $ \Phi $ and $ \Phi^{\dagger}$. In the free field 
 limit, the construction of composite operators such as 
 $ \chi_R ( \Phi ) $ is simple. No short distance 
 subtractions are required, due to  the vanishing  two point function 
\bea 
 < \Phi ( x ) \Phi ( y ) > = 0 
\eea 

 If  we wish to consider a local operator 
 of the form  $ tr ( \Phi ) tr  ( \Phi^{\dagger} )  $ we have 
 to subtract a  short distance singularity. Denoting the well-defined 
 local operator as  $ : tr ( \Phi ( x)  ) tr  ( \Phi^{\dagger} ( x ) ) :  $
 we have
\bea 
 : tr ( \Phi ( x)  ) tr  ( \Phi^{\dagger} ( x ) ) : 
 ~ = Lim_{ \epsilon \rightarrow 0 } ~~ 
  tr ( \Phi ( x)  ) ~ tr  ( \Phi^{\dagger} ( x + \epsilon  ) )
  -  { N  \over \epsilon^2 }     
\eea 
Note that the renormalised operator leads to a well-defined 
 state 
\begin{equation}
 :   tr ( \Phi ( x)  ) tr  ( \Phi^{\dagger} ( x ) ) :| 0 >  \nnm 
\end{equation}
For example in computing the overlap of this state with $ < 0 | $
we get a well-defined correlator. Without the subtraction we 
would get a divergent answer for this overlap.

 A naive guess for the gauge theory dual of 
 a brane-anti-brane system would be 
 $  : ~\chi_R ( \Phi ) \chi_S ( \Phi^ { \dagger} )~  :  ~~  $.  
 Such an operator is not in general orthogonal to 
 operators which are of the form  
$  : tr ( \Phi \Phi^{\dagger} ) \chi_{R_1}
  ( \Phi  ) \chi_{S_1} ( \Phi^{\dagger} ) : ~ $.
 In the simplest case of $m =1 , n =1 $, for example, 
  $  : tr  ( \Phi )  tr  ( \Phi^{\dagger} ) : $ is not orthogonal 
 to $ : tr ( \Phi \Phi^{\dagger} ): $ ~ .  
 However consider, 
in this case, an operator 
\bea 
\cO  = tr ( \Phi ) tr ( \Phi^{\dagger} ) -  { 1 \over N }
 tr ( \Phi \Phi^{\dagger} ) 
\eea 
This has a number of interesting and 
 easily verified properties. The first is that 
its short distance subtractions vanish 
\bea 
: \cO : ~  = ~ \cO
\eea 
The second is that 
\bea 
< \cO   ~  :  tr ( \Phi \Phi^{\dagger } ):  > = 0 
\eea

In terms of the Young diagram classification of operators, 
$ tr ( \Phi ) $ corresponds to the single box Young diagram
denoted as $ [1] $. 
So the operator $ \cO$ above can be viewed as a
singularity free operator associated to 
the pair of Young diagrams $ ( [1] , [1] ) $, which is 
orthogonal to operators where $ \Phi , \Phi^{\dagger} $ 
are in the same trace.  
Brauer algebras will allow us to associate such a 
singularity free operator to any pair of Young diagrams 
$ R , S $ in the large $N$ limit. More precisely we will 
need $c_1  ( R ) + c_1 ( S ) \le N $, where $c_1$ denotes the 
length of the first column of the Young diagram. 
 The origin of this condition in representation theory comes from  
(\ref{nonchirschwe}). When   we consider a brane 
$R$ and antibrane $S$ with   $c_1  ( R ) + c_1 ( S )  >  N $, 
their composite is best viewed as an excited state of another pair 
of branes satisfying the bound. 
A related fact is that the naive guess 
$ : \chi_R ( \Phi ) \chi_S ( \Phi^{\dagger} )  : $ 
becomes completely dependent on operators where 
 $ \Phi , \Phi^{\dagger} $ appear in the same trace.
This will be explained in section 8. 

The Brauer technology is also useful in classifying operators 
in a zero-dimensional or one-dimensional Matrix Model.
In the one-dimensional case it has been shown \cite{cjr} 
(see also  \cite{taktsu,djr,rodrigues}) that the reduction of the 
four-dimensional action on  $S^3 \times R $ leads to 
the Hamiltonian and $SO(2) $ symmetry generator  
\bea 
&& H = tr  ( A^{\dagger } A + B^{\dagger } B )  \cr 
&& J = tr  (  A^{\dagger } A - B^{\dagger } B ) 
\eea 
The construction of operators corresponding to giant gravitons 
involves gauge invariant states obtained by acting with 
$ A^{\dagger} $ on the vacuum. For anti-giant gravitons, 
we act with $B^{\dagger} $ only. For systems consisting of 
composites of giant and anti-giant we act with both $A^{\dagger} $ and
 $ B^{\dagger} $. We will find a diagonal basis in the space of 
such operators using the Brauer algebra. 
The two-point function (\ref{bas2pt}),  with its 
position dependence removed,  also appears in the 
zero-dimensional complex matrix model  introduced 
in \cite{ginibre} and studied more recently in 
\cite{kostau,kostau2,kpss}, which has a  partition function
\bea\label{zeropart}  
  Z = \int [ d \Phi d \Phi^{\dagger} ]  
e^{ - tr ( \Phi \Phi^{\dagger}  ) } 
\eea
 Hence our results are also relevant 
to this zero-dimensional matrix model. Finally we expect that 
the results on projectors should help in a better understanding 
of the stringy interpretation of the non-chiral large $N$ expansion 
of intersecting Wilson loops in two-dimensional Yang-Mills (2dYM).  
We will be making contact with some of the character
and dimension formulae which play a role in the non-chiral 
expansion of 2dYM \cite{grotay,grotay2,cmr}.

Having given away what Brauer algebras do for us,  
it is time to describe them more precisely and 
to explain why they are useful in understanding 
the properties of large composite operators 
in gauge theory.

\section{Gauge invariant operators, correlators and Brauer Algebras } 

In understanding the role of Brauer algebras in 
the calculation of correlation functions of operators,  it is 
useful to express the basic formulae from field theory 
in a diagrammatic notation for operators acting on 
$ V $ or $ V^{\otimes n } $ or more generally 
on $  V^{\otimes m } \otimes V^{* \otimes  n } $. 
The use of diagrams for representing operators 
on tensor space plays a crucial role in knot theory \cite{reshtura} 
and has also been developed by physicists \cite{cvi}. 
In \cite{srwil} the tensor space diagrams are used 
in the calculation of Wilson loops in two-dimensional 
Yang-Mills theory. In section 3 of \cite{cr}
  some of these diagrammatic techniques 
were summarised and  used to simplify proofs of properties of 
correlators of large dimension multi-traces \cite{cjr}.

\subsection{ Correlators and Brauer algebras }

\begin{figure}[t]
\begin{center}
 \resizebox{!}{4cm}{\includegraphics{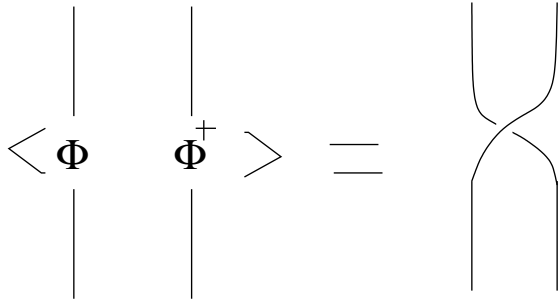}}
\caption{ Correlator of $ \Phi $ and $ \Phi^{\dagger} $ 
as permutation operator } \label{fig:phiphidagcor}
\end{center}
\end{figure}

\begin{figure}
\begin{center}
 \resizebox{!}{4cm}{\includegraphics{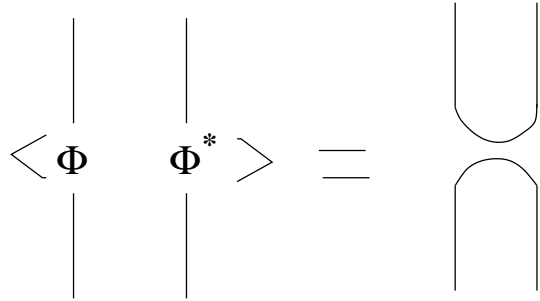}}
\caption{ Correlator of $ \Phi $ and $ \Phi^* $ as contraction operator  }
 \label{fig:phiphistacor}
\end{center}
\end{figure}

\begin{figure}
\begin{center}
 \resizebox{!}{4cm}{\includegraphics{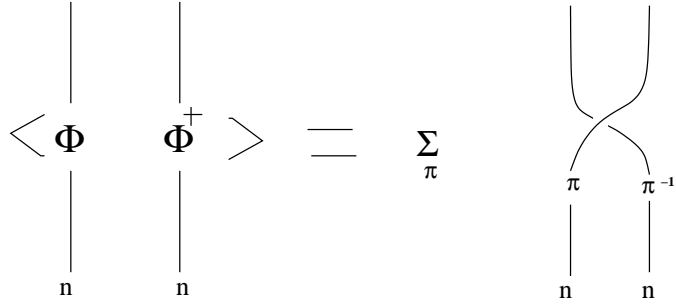}}
\caption{ Correlator of $n$ copies of  $ \Phi $ and  $n$
copies of  $ \Phi^{\dagger} $   }
 \label{fig:phiphidagcorn}
\end{center}
\end{figure}

Consider  the basic 2-point function obtained by  doing the free-field 
path integral over Matrices
\bea\label{2ptindex}
\langle \Phi^i_j \Phi^{\dagger k }_{l} \rangle = \delta^i_l \delta^k_j 
\eea 
where we have dropped the spacetime-dependence. 
By recognising $ \Phi $ as an operator on $ V $ and  
 $ \sigma $ as an operator on $ V \otimes V$
with matrix elements 
\bea 
  (\sigma)^{ik}_{jl}  =  \delta^i_l \delta^k_j
\eea 
we can re-write (\ref{2ptindex}) in an index-free form as 
\bea\label{indfree2pt}  
\langle \Phi \otimes \Phi^{\dagger}  \rangle = \sigma 
\eea
This can be expressed in a precise diagrammatic form in 
Figure \ref{fig:phiphidagcor}. The permutation on the RHS  is
 a map from $ V \otimes V $ to $ V \otimes V $. 
The power of the diagrammatic presentation comes from the fact 
that essentially the same diagram represents the 
2-point function when we have $n$ copies of $\Phi $ 
and $n$ copies of $\Phi^{\dagger}$. Now we have $\Phi $ 
as an operator on $V^{\otimes n } $ which is simply denoted
by a line labelled by $n$. $ \Phi $ is understood to act on this 
as $ \Phi \otimes \Phi \otimes  \cdots  \otimes \Phi $. The result of the 
correlator is to have the same twist, but in addition, 
a sum over permutations in $ S_n $  (denoted by $\pi$ in Figure 
\ref{fig:phiphidagcorn}) which determines which 
of the $n$  $ \Phi$'s is contracted with which of the $n$
$\Phi^{\dagger} $ (for uses of this diagrammatic formula 
see \cite{cr}).

Rather than writing the 2-point function in terms of 
$ \Phi $ and $ \Phi^{\dagger} $ we can use the complex conjugate 
$ \Phi^* $ to write 
\bea\label{2ptstarindex}  
\langle \Phi^i_j \Phi^{* k }_{l} \rangle = \delta^{ik} \delta_{jl}  
\eea 
We can view $ \Phi^* $ as a map from conjugate  $\bar  V $ to $\bar V $ 
and $ \Phi \otimes \Phi^* $ as a map from 
$ V \otimes \bar V $ to $ V \otimes \bar V $. To describe the 
right-hand side of (\ref{2ptstarindex}) in algebraic terms we introduce 
the linear operator  $ C  $ as a map  from  $ V \otimes \bar V$
to  $ V \otimes \bar V $
\bea 
  (C)^{ik}_{jl}  = \delta^{ik}  \delta_{jl}  
\eea 
 We can now write the index free form of 
(\ref{2ptstarindex}) as 
\bea \label{indfree2pt2} 
\langle \Phi \otimes \Phi^{*}  \rangle = C 
\eea
This can be expressed in a precise diagrammatic form in 
 Figure \ref{fig:phiphistacor}. 

Note that in (\ref{2ptstarindex})
 the indices  $k,l $ are viewed as labelling 
vectors in $\bar V$ whereas the $(i,j)$ denote vectors in 
$ V $.  We can make that explicit in the diagrams by introducing 
arrows, but this is a refinement of  the diagrammatic notation, 
which is not  crucial, though it is sometimes useful 
in manipulating the diagrams.

\subsection{  Brauer Algebra : definition and Schur-Weyl duality } 

We review known facts about Brauer algebras mostly from  \cite{halverson}. 
Other useful references are  \cite{stem, koike, ramthesis, bch,ram,naza}. 

Recall that $ S_n  $ is the centraliser of 
$U(N)$ ( or $GL(N)$ )  acting on $ V^{\otimes m } $. 
Hence we have Schur-Weyl duality
\bea 
V^{\otimes n } = \oplus_{ R  } V_R^{ U(N) }  \otimes V_R^{ S_n } 
\eea 
$S_m \times S_n$ is contained in  centraliser 
of $U(N)$ ( $GL(N)$ ) 
 acting in  $ V^{\otimes  m } \otimes \bar V^{\otimes  n } $. 
But we also need contractions, which along with 
the permutations, generate the algebra $ B_N ( m  , n ) $. 
Hence Schur-Weyl duality  states that 
\bea\label{nonchirschwe}  
V^{  \otimes m } \otimes \bar V^{ \otimes n }
= \oplus_{ \gamma } V_{ \gamma }^{   U(N) } \otimes 
V_{\gamma } ^{ B_N (m,n)   } 
\eea 
It gives the decomposition of the tensor product $\vmn$ 
in terms of irreps of $U(N)$ and $B_N ( m,n)$.   
$ \gamma $ runs over  sets of integers 
$ ( \gamma_1 , \gamma_2 , \cdots , \gamma_N )$ obeying
$ \gamma_1 \ge \gamma_2 \ge \cdots \ge \gamma_N $.
The set of positive integers defines $ \gamma_+$ which is a 
partition of $ m - k $  while the 
negative integers define a partition $ \gamma_{-} $ of 
$ n - k$. Here $k$ is an integer lying between $0$ and $min ( m,n)$. 
 Equivalently $ \gamma_+ $  determines a Young diagram 
with $m-k$ boxes, $\gamma_- $ one of $n-k$ boxes.   A choice of  $ \gamma $ 
 is equivalent to a choice of $ ( k , \gamma_+ , \gamma_- ) $. 
If we write $\gamma_+ $ as a Young diagram, with row lengths 
equal to the parts in the partition, $c_1 ( \gamma_+ ) $ is defined 
as the length of the first column.   It follows from the above 
definitions  that $ c_1 ( \gamma_+ ) + c_1 ( \gamma_- ) \le N $. 
See more details on this in section  \ref{sec:finiteN}.

\begin{figure}[t]
\begin{center}
 \resizebox{!}{4cm}{\includegraphics{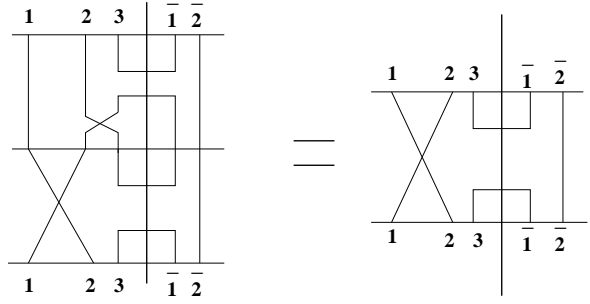}}
\caption{Example of product in Brauer algebra  }
 \label{fig:brauerprodexamp}
\end{center}
\end{figure}

\begin{figure}[t]
\begin{center}
 \resizebox{!}{4cm}{\includegraphics{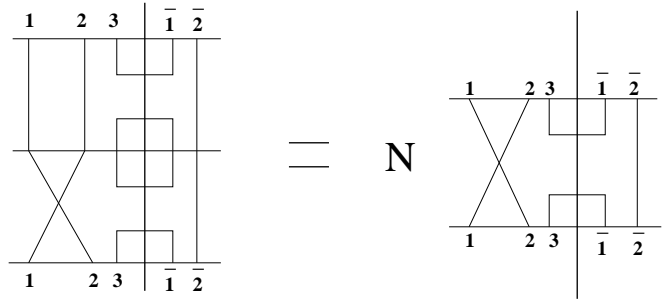}}
\caption{Example of product in Brauer algebra with loop giving $N$   }
 \label{fig:brauerprodexamp1}
\end{center}
\end{figure}

From the definition of the Brauer algebra elements in terms 
of operators in tensor space, we can derive diagrammatic rules 
for multiplying them. The multiplication is done by stacking the 
diagrams corresponding to the Brauer elements. 
Symmetric group elements in $S_n$  can be represented
diagrammatically using   two horizontal lines each 
containing $n$ marked  points labelled 
by integers $1 ... n $. We will refer to these as two rungs. 
Any particular element  $ \sigma \in S_n $ is represented 
by drawing lines joining an integer $i$ from the bottom 
rung to an integer  $\sigma(i)$ in  the top rung. 
Multiplication of elements in $S_n $ is obtained by 
stacking one pair of rungs on top of another, 
and identifying the top of the bottom pair 
to the bottom of the top pair. 
Elements in $S_m \times S_n$ are represented using 
two rungs with integers $1 ... m $ on the left  side of a vertical 
barrier and $  \bar 1 ..  \bar n $ on the right side of the vertical barrier. 
Brauer elements in $ B_N  ( m , n ) $ are drawn using two 
horizontal rungs as for $S_m \times S_n $, but now in addition 
to the  lines of $ S_m \times S_n$ we allow 
lines joining the points on the  lower (upper) left of the barrier 
to points on the lower (upper) right of the barrier. 
Multiplication is done as before by stacking two pairs 
of rungs. Closed loops are replaced by the parameter 
$N$. This is illustrated in Figures \ref{fig:brauerprodexamp}, 
\ref{fig:brauerprodexamp1}. In Figure \ref{fig:brauerprodexamp}
we have 
\bea 
  (  C_{3 \bar 1 } (23)  ) \cdot ( C_{3 \bar 1 } (12) ) = C_{3 \bar 1 } (12)
\nonumber
\eea
In Figure \ref{fig:brauerprodexamp1} we have 
\bea 
  (  C_{3 \bar 1 } ) \cdot ( C_{3 \bar 1 } (12) ) = N  C_{3 \bar 1 } (12)  
\nonumber
\eea

The Brauer algebra can be described by generators and 
relations. The relations can be obtained by diagrammatic 
manipulation. The generators include the 
simple transpositions $s_i$ of $ S_m $ and 
$ \bar s_i $ of $S_n$. The simple transposition 
$s_i$ exchanges $i$ with $i+1$ leaving everything else fixed.
To this we add $C_{1\bar 1 } $ which contracts the first 
$V$ factor with the first $ \bar V $ factor.  By using the diagrammatic 
approach, one easily derives relations such as   
\bea\label{somerels} 
&& C_{i \bar j } = (i1) (\bar 1 \bar j ) C_{1 \bar 1 }  (i1) (\bar 1 \bar j ) 
\cr 
&&  C_{ i \bar j } (ik) C_{i \bar j } = C_{ i \bar j } \cr 
&& C_{i \bar j  }   C_{ i \bar k } = C_{i \bar j } ( \bar j \bar k  ) 
                     = ( \bar j \bar k ) C_{i \bar k } \cr 
&& C_{i \bar j }  C_{ k \bar j } = C_{i \bar j } ( ik ) = (ik)   C_{ k \bar j }
 \eea 
Of course these can also be derived equivalently by 
writing out   all the operators 
involved in terms of their matrix elements in
 $ V^{ \otimes m } \otimes \bar V^{\otimes n } $. 

It is also easy to check that Brauer elements commute with 
the action of the Lie algebra of 
 $ GL(N) $ or $U(N)$ in $ V \otimes \bar V $. 
Let $E_{ij}$ be the matrix with $1$ in the $(i,j)$ entry, 
and $0$ everywhere else. 
We have 
\bea 
&& ( E_{ij} \otimes 1 + 1 \otimes E_{ij } ) ~ C ~ 
  v_{k} \otimes \bar v_{l } \cr 
&& = ( E_{ij} \otimes 1 + 1 \otimes E_{ij } ) ~ \delta_{kl} ~  v_m \otimes 
\bar v_m \cr 
&& = \delta_{kl} ( \delta_{jm} v_i \otimes \bar v_m - \delta_{im} v_m \otimes \bar v_j ) \cr 
&& = 0 
\eea 
and 
\bea 
&& C ( E_{ij} \otimes 1 + 1 \otimes E_{ij } )  v_{k} \otimes \bar v_{l } \cr 
&& = C ( \delta_{jk} v_i \otimes \bar v_{l}  - \delta_{il} 
v_k \otimes \bar v_{j } ) \cr 
&& = \delta_{jk} \delta_{il} v_m \otimes \bar v_m - \delta_{il}\delta_{jk} v_m \otimes \bar v_m ) \cr 
&& =0 
\eea

We record here the formula for dimensions of Brauer representations,  
\bea\label{brauerdim}  
 d^{(B)}_{ \gamma } =   { m! n! \over k! ( m-k)!  (n -k ) ! }
 {  d_{\gamma_+ } d_{ \gamma_-}   }  
\eea 
in terms of the dimensions $d_{\gamma_+} $ of the $ S_{m-k} $ 
representation associated with the partition $\gamma_+ $, 
and $  d_{\gamma_-} $ of the $ S_{n-k} $ 
representation associated with the partition $\gamma_- $. 
There is also a useful formula for the multiplicity of an irrep 
$ ( \alpha , \beta ) $ of  $ \mathbb  C  [ S_m \times S_n ] $
subalgebra of $B_N(m , n ) $ appearing in the irrep 
$ \gamma $ of the Brauer algebra 
\bea\label{redcoeff}  
M^{\gamma }_{ ( \alpha , \beta ) } 
= 
  \sum_{\delta \vdash k } g ( \delta , \gamma_+ ; \alpha ) 
                          g ( \delta , \gamma_- ; \beta ) 
\eea 
Here 
$\delta \vdash k$ 
expresses the fact that $ \delta $ is a partition of $k $. 
$g ( \delta , \gamma_+ ; \alpha )$ is the Littlewood-Richardson (LR) 
coefficient, determined by putting together Young diagrams
according to certain rules (see for example \cite{fulhar}). 
We will often use a single label $A$ for the irreps 
of  $ \mathbb  C  [ S_m \times S_n ] $ where it is understood 
that $A =  ( \alpha \vdash m , \beta \vdash n )$, so that the 
multiplicity is written as $M^{\gamma}_A $.

\subsection{ The bilinear form and map between $ B_N  ( m,n )$ to 
$\mathbb C [S_{m+n}]$  } 
\label{sec:mapBCS}

There is a symmetric bilinear form on the Brauer algebra, which follows 
by viewing them as operators in tensor space 
\bea\label{bilfor}  
< b_1 , b_2 > = tr_{m,n}  ( b_1 b_2 ) 
\eea 
Above and in the rest of the paper $tr_{m,n} $ denotes the trace 
 taken in $ V^{\otimes m } \otimes \bar V^{ \otimes n } $. 
Using the  bilinear form we can define the dual element $b^*$ 
of any element $b$ by the property 
\bea 
tr_{m,n}  ( b b^* ) =1 
\eea 
It is shown in \cite{ramthesis} that for any fixed element $c$ 
the following sum 
\bea 
[c] = \sum_{b} b c b^* 
\eea 
 over a complete basis of $B_N(m,n)$, gives 
a central element, which commutes with any  $ b \in B_N  ( m , n ) $. 
This is a generalisation to semi-simple algebras of 
the group averaging procedure for group algebras. 
The dual elements also allow a construction of projectors 
\bea\label{gencentproj}  
P^{ \gamma }
&=&
t_{\gamma }\sum_{b}\chi_{\gamma }(b)b^{\ast} =  
t_{\gamma }\sum_{b}\chi_{\gamma }(b^{\ast} )b 
\eea
where $b$ runs over a basis for $B_N (m,n)$.  
The normalisation factor can be seen, in this case,  to be 
 $ t_{\gamma } =  { Dim \gamma } $, the dimension of 
the $U(N)$ irrep associated with the label $ \gamma $. 
 This  follows from \cite{ramthesis} where it is shown 
$P^{\gamma}$  is a central projector 
(idempotent), with $ t_{\gamma } $ equal to the trace of a matrix unit 
$ tr ( E^{\lambda }_{11} ) $. 
In our case the trace is being taken in 
$ V^{\otimes m } \otimes \bar V^{\otimes n } $  and using Schur-Weyl duality
(\ref{nonchirschwe}) we find that 
\bea 
 t_{\gamma} &=& \sum_{ \lambda }  ( Dim \lambda  )~  tr_{\lambda} 
(  E^{\gamma }_{11} )  \cr 
& =& \sum_{ \lambda }  ( Dim \lambda ) ~ \delta_{\lambda \gamma}  \cr 
& =& Dim \gamma 
\eea

In the case at hand,  we can obtain a lot of information about the 
bilinear form (\ref{bilfor}) by exploiting a map $ \Sigma $
\bea 
\Sigma : B_N ( m , n ) \rightarrow \mathbb C [S_{m+n } ]
\eea

\begin{figure}[t]
\begin{center}
 \resizebox{!}{3cm}{\includegraphics{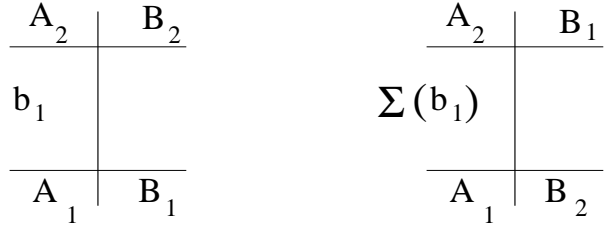}}
\caption{The map $\Sigma $ from $ B_N (m,n) $ to 
$\mathbb C [S_{m+n}]$  } \label{fig:sigmamap}
\end{center}
\end{figure}

\begin{figure}
\begin{center}
 \resizebox{!}{6cm}{\includegraphics{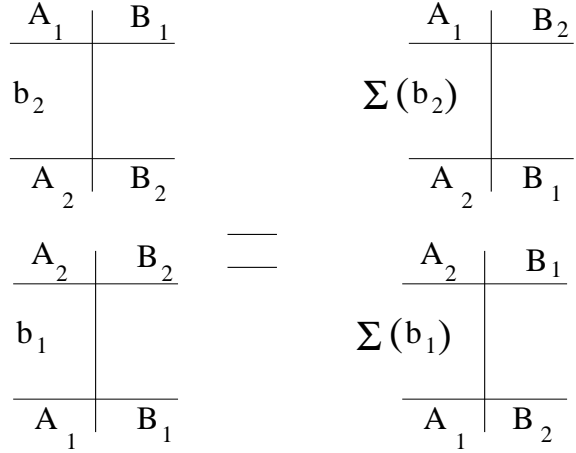}} 
\caption{Showing that $tr (b_1b_2 ) = tr ( \Sigma(b_1) \Sigma(b_2) )$  } 
\label{fig:trbbistrsbsb}
\end{center}
\end{figure}

We recall that in   the diagrammatic description of Brauer elements 
given earlier in this section, we have two horizontal lines, 
one on top of the other. Each line has $m+n$ points, with a vertical 
barrier separating the $m$ from the $n$. In $B_N(m,n)$ lines crossing 
the barrier join points on the upper left to points on the 
upper right, and points on the lower left points on the lower right. 
Elements on $ S_{m+n}$ are described by lines joining 
lower points to upper points whether they cross the barrier or not. 
The map $\Sigma $ simply reflects the upper right segment 
and  the lower right segment into each other.  
It is illustrated in Figure \ref{fig:sigmamap}, 
where $ B_1 , B_2  $ denote the sets of points labelled $ 1.. n $. 

It is an invertible map from 
$  B_N(m,n) $ to $ \mathbb C [S_{ m + n }] $, which is consistent 
with the fact that the dimension of the Brauer algebra as a vector space is known to be $ (m+n)!$ ( e.g see \cite{halverson} ).   
 It is not a homomorphism. 
It however has the crucial property that it maps the 
symmetric bilinear form on $B_N(m,n)$ to a symmetric bilinear form 
on  $ \mathbb { C } [S_{m+n }] $  
\bea\label{mapbilfor}  
\fbox{
$\displaystyle{
  tr_{m,n} (b_i b_j ) = tr_{m,n}   ( \Sigma (b_i ) \Sigma ( b_j ) ) 
}$
}
\eea
where $b_i$ runs over a complete basis for the Brauer algebra. 
This is made clear by  Figure \ref{fig:trbbistrsbsb}. 
On the left hand side, the set of points labelled by $B_2 $ 
(a set of labels for tensor space indices) on  the upper diagram for 
$b_2$ is identified with the set $B_2$ on the lower diagram for 
$b_1$ by the multiplication $b_2b_1$. The sets labelled by 
 $B_1$ are identified  by the trace. On the right hand side, the $\Sigma $ map 
performs the reflection of labels $ B_1 \leftrightarrow B_2 $. 
The multiplication of $ \Sigma (b_2) $ with $ \Sigma(b_1)$ 
identifies the $B_1$ sets. The trace identifies the $B_2$ sets.  
The outcome in both cases is determined by the $B_1, B_2$ identifications, 
which proves (\ref{mapbilfor}). 
An explicit formula  for the bilinear form follows
\bea 
  tr  ( \Sigma (b_i ) \Sigma ( b_j ) ) 
&&   = \sum_{ T \vdash m +n } Dim T ~ \chi_T (  \Sigma (b_i ) \Sigma (b_j) ) 
  \cr 
&& =    N^{m+n} \delta  ( \Omega_{m+n} \Sigma (b_i ) \Sigma (b_j) )
 \eea 
The delta function over the symmetric group algebra 
$\delta ( \sigma ) $  is defined to be $1$ if $\sigma $ is the identity 
and $0$ otherwise.   
For $ m + n \ge N $, 
the  sum over $T$ is restricted by the condition $ c_1 ( T ) \le N $. 
The $\Omega_{m+n}$ factor is familiar from 2dYM theory,
and is defined by  
\bea 
\Omega_{n  } =  \sum_{ \sigma \in S_{ n } } N^{C_{\sigma} - n } \sigma 
\eea 
where $ C_{\sigma } $ is the number of cycles in the permutation 
$ \sigma $. When $ n < N $ ,  $ \Omega_n $ can be inverted,
 and this is used to good effect in the large $N$ expansion of 2dYM.  
Here we have 
\bea 
(\Sigma (b) )^* = N^{-m-n} \Omega_{m+n}^{-1}   ( \Sigma (b ) )^{-1} 
\eea 
Since  $ \Sigma $ preserves the bilinear form  (\ref{mapbilfor}) we have 
$( \Sigma  (b) )^* = \Sigma (b^* ) $ 
\bea\label{bstar}  
b^* =   N^{-m-n}   \Sigma^{-1}  
( \Omega_{m+n}^{-1}   ( \Sigma (b ) )^{-1} )
\eea 
An expansion of the $\Omega_{m+n}^{-1} $ in terms of 
characters can be given 
\bea\label{omeginv}  
 \Omega_{m+n}^{-1} &&= {  N^{m+n} \over ( m+n )! }  \sum_T 
 { d_T \over Dim T } p_T \cr 
&& =  {  N^{m+n} \over  (( m+n )!)^2 }  \sum_T 
 { d_T^2 \over Dim T } \sum_{ \sigma \in S_{m+n} } \chi_T ( \sigma ) ~ \sigma 
\eea
so that
\bea 
(\Sigma (b ))^* =  { 1 \over ( ( m + n ) !)^2 }
\sum_{T } {d_T^2 \over Dim T } \sum_{ \sigma \in S_{m+n } } 
  \chi_T ( \sigma ) \sigma ( \Sigma (b ) )^{-1} 
\eea 
and 
\bea\label{bstar1}  
\fbox{
$\displaystyle{
b^* = { 1 \over ( ( m + n ) !)^2 } \sum_{T } 
{ d_T^{2} \over Dim T }  \sum_{\sigma \in S_{m+n } } 
 ~ \chi_T ( \sigma  ) ~  \Sigma^{-1} (~  \sigma ( \Sigma ( b ))^{-1} ~  ) 
}$
}
\eea 
For $m + n \le  N $ the sum $T$  runs over all partitions 
of $m+n $, but more generally $c_1 ( T ) \le N $.

It is instructive to consider 
\begin{equation}
G_{ij}=tr_{m,n} (b_{i}b_{j})=tr_{m,n}  (\sigma_{i}\sigma_{j})
\end{equation}
where $b_{i}$ belongs to a basis set  in $B_N(m,n)$ while 
$\sigma_{i}$ is the corresponding element $ \Sigma ( b_i ) $ in $S_{m+n}$. 
The inverse  is defined as
\begin{equation}
\sigma^{\ast}_{i}=G^{ij}\sigma_{j}
\end{equation}
From this equation, we obtain
\begin{equation}
\delta(\sigma^{\ast}_{i}\sigma_{k}^{-1})
=G^{ij}\delta(\sigma_{j}\sigma_{k}^{-1})=G^{ik} 
\end{equation}
\begin{eqnarray}
G^{ij}
&=&
\delta(\sigma^{\ast}_{i}\sigma_{j}^{-1}) \cr
&=&
\frac{1}{N^{m+n}}\delta\left(\Omega_{m+n}^{-1}
\sigma_{i}^{-1}\sigma_{j}^{-1}\right) \cr
&=&
\frac{1}{\left((m+n)!\right)^{2}}
\sum_{ T }\frac{d_{ T }^{2}}{dim T }\chi_{ T }
(\sigma_{i}^{-1}\sigma_{j}^{-1})
\end{eqnarray}
As usual at finite $N$ we restrict $ c_1 ( T ) \le N $.
We have used the following equations :  
\begin{eqnarray}
&&\delta(\sigma)=\frac{1}{(m+n)!}\sum_{ T }d_{ T }\chi_{ T }(\sigma) \cr
&&dim T =\frac{N^{m+n}}{(m+n)!}\chi_{ T }(\Omega_{m+n})
\end{eqnarray}

\subsection{ Brauer algebra as a spectrum generating algebra 
 for multi-traces of two Matrices  }\label{specgen} 

In the case of holomorphic multi-trace operators constructed from 
$n$ copies of $ \Phi $ it was useful to write them 
as $ tr_n ( \sigma \Phi ) $, where the trace is in $V^{\otimes n } $
\cite{cjr}.  
Different multi-trace operators correspond to different 
states in the Hilbert space of $ N=4 $ SYM on $ S^3 \times R $, 
via the operator-state correspondence (for elaboration of this see 
 \cite{witten,pp,copto}). When two permutations $ \sigma_1 , \sigma_2 $ 
in $S_n $ are related by a permutation $ \sigma_3 $ as 
$ \sigma_1 = \sigma_3 \sigma_2 \sigma_3^{-1} $, they lead to 
the same multi-trace operator, and the same state in the Hilbert space 
in $ N=4 $ SYM. Permutations $ \sigma \in S_n $, subject 
to an equivalence relation of conjugation by another element 
in $S_n $, are just the conjugacy classes of $S_n$. In each equivalence 
class we have a central element :  the sum of permutations 
in the conjugacy class which is  proportional to the group average 
$ \sum_{ \sigma_1 } \sigma_1 \sigma \sigma_1^{-1} $.  The conjugacy classes 
of $S_n $ can be viewed, therefore, as forming a  spectrum generating algebra.

In the non-chiral case at hand,  where we are considering 
multi-traces constructed from  $ \Phi , \Phi^{\dagger} $, 
the Brauer algebra plays an analogous role. In the simplest 
case of $B_N(1,1)$, the Brauer algebra is two-dimensional, spanned 
by $1$ and $ C$. We have
\bea 
&&  tr_{(1,1)}  ( \Phi \otimes \Phi^* )  = tr \Phi  tr \Phi^{\dagger}      
\eea 
and 
\bea 
&&  tr_{(1,1)}  C ( \Phi \otimes \Phi^* )\cr 
&& = < e^i \otimes \bar e^j | C ( \Phi \otimes \Phi^* ) 
 | e_i \otimes \bar e_j > \cr 
&& =  < e^i \otimes \bar e^j  | C \Phi_i^k \Phi^{ *  l }_{j} 
| e_k \otimes e_l >  \cr 
&& =  < e^i \otimes \bar e^j  | \delta_{kl } \Phi_i^k \Phi^{ *  l }_{j} | e_m \otimes \bar e_m > \cr 
&& = tr \Phi \Phi^{\dagger} 
\eea 
In fact  any multi-trace constructed from 
 $ \Phi , \Phi^{\dagger}  $ can be obtained by using 
$ tr ( b ~  \Phi \otimes \Phi^{\dagger} ) $ for  general $b$. 
Consider any trace $ tr_{m,n}  ( \Phi^{m_1} \Phi^{\dagger  ~ n_1 } 
 \Phi^{m_2} \Phi^{ \dagger  ~ n_2 } \cdots \Phi^{m_k} 
 \Phi^{\dagger  ~ n_k }   ) $. 
 The element $b$ 
in this case involves cyclic permutations with cycle lengths 
$m_1, m_2 ... $ in $S_m $ and a permutation with cycles $ n_1, n_2  ... $
in $ S_n $ along with $k$ contractions and  a further 
permutation to join the contractions.  One can 
write an explicit formula to demonstrate the above, 
but  the reader should easily convince him(her)self 
of the above claim by considering a few examples. 
An important point to note is that a Brauer element $b$ 
and another element $ h  b h ^{-1} $ produce the same multi-trace 
if $h \in S_m \times S_n $. In the diagrammatic presentation of 
Brauer elements, the conjugation corresponds to re-labelling 
the numbers on the top and bottom rungs. 
Therefore,  counting  multi-traces is the same as counting these equivalence
 classes under conjugation by $h$. 
In each equivalence class we can build, 
 by averaging any chosen element $b$ using 
$ \sum_h h b h^{-1} $, a unique element which commutes with $ S_m \times S_n$. 
We will call such elements, {\bf symmetric  elements}. 
It is interesting that this notion of 
equivalence by conjugation with $h\in S_{m}\times S_{n}$ 
has been studied in \cite{halverson} purely as 
an algebraic property. 
Here the equivalence is motivated by the role of Brauer as a 
spectrum generating algebra for multi-traces. 

A natural way to construct symmetric elements from representation 
theory is to consider projectors for fixed irreducible representations. 
In sections 4, 5 we will be considering projectors for
Brauer irreps. Since each Brauer irrep decomposes under the 
action of the $ \mathbb C [ S_m \times S_n ] $ sub-algebra 
into irreducible reps according to (\ref{redcoeff}), the 
central Brauer projectors will be sums of {\it symmetric Brauer  projectors}. 
However the symmetric Brauer projectors do not exhaust the 
complete set of symmetric elements. One has to consider more general 
{\bf symmetric branching operators}. We will describe them in more 
detail in section 7, where we will also argue that they provide 
a complete set of symmetric elements. In section 6, we will 
show how to count symmetric projectors, and symmetric branching 
operators, and show agreement with the known counting of large $N$ 
multi-traces based on Polya theory. The Brauer counting however extends beyond 
 the large $N$ limit, and finite $N$ effects will be described in 
section \ref{sec:finiteN}. 

We observe some useful facts,  which involve the map 
$ \Sigma $ given in section \ref{sec:mapBCS}. 
The map $ \Sigma $ is also useful when we construct 
gauge invariant operators from the matrices $ \Phi , \Phi^{\dagger } $.
The equation 
\bea\label{sigpresmet}  
 tr_{m,n} ( b \Phi \otimes \Phi^* )  = 
tr_{m,n} ( \Sigma ( b ) \Phi \otimes \Phi^{\dagger}  )
\eea 
follows from the diagrammatics.
In the Matrix quantum mechanics we have the  related fact that 
\bea 
 tr_{m,n} (\Sigma (  b )  A^{\dagger}  \otimes B^{\dagger}  )  = 
tr _{m,n}(  b  A^{\dagger}  \otimes  ( B^{\dagger} )^T   )
\eea 
The transpose acts on the matrix indices. It is also useful to note 
that 
\bea\label{symsym}  
h b h^{-1}    = b  \quad   \Longleftrightarrow \quad h \Sigma ( b ) h^{-1} = \Sigma ( b ) 
\eea 
In other words, if $b$ is a symmetric element, then $ \Sigma (b) $ is 
a symmetric element.

\section{ Projectors in Brauer algebra }
\label{sec:projectorinBrauer}

As reviewed in the discussion of (\ref{nonchirschwe}) 
the tensor product $ 
V^{\otimes m } \otimes \bar V^{\otimes  n }$ 
decomposes into a direct sum of irreps of $B_N(m,n)$ and  $U(N)$ 
labelled by $ \gamma$, where  $ \gamma = ( k , \gamma_+ , \gamma_- ) $. 
For each $ \gamma $, there is a projector $ P^{\gamma } $ which, 
acting on $ \vmn $,  projects onto the subspace labelled 
by $\gamma $. This projector can be constructed from 
Brauer elements, and is central, i.e commutes with any element of 
$B_N ( m,n )$.  The $k=0$ projectors have special properties. 
Writing $ P^{ (k=0, \gamma_+ = R ,\gamma_- = S)  } \equiv P_{R \bar S } $ 
to connect to the notation of 2D Yang Mills theory, we have, 
 for a unitary matrix $ U$,  
\bea\label{projcharnc}  
tr_{m,n} ( P_{ R \bar S } U )  = d^{(B)}_{ R \bar S }
\chi_{R \bar S } ( U )    
               = d_R d_S \chi_{R \bar S } ( U )  
\eea 
On $\vmn$,  
$U$ acts as $ U \otimes U \cdots U \otimes U^* \otimes \cdots U^* $ 
with $m$ factors of $U$ and $n$ factors of $U^*$.  
The first equality follows from (\ref{nonchirschwe}) and the second equality 
expresses the Brauer dimension using (\ref{brauerdim}) in terms of 
the symmetric group irrep dimensions $d_R , d_S $. The character 
$ \chi_{R \bar S } ( U ) $ is the character of the ``coupled representation''
used in 2dYM \cite{grotay,grotay2}.  By setting $U=1$ we have 
\bea\label{tpdimf}  
tr_{m,n} ( P_{ R \bar S }  ) = d_R d_S Dim R \bar S
\eea  
In section \ref{sec:orthogonalsetoperator}
 we will be considering local operators in 4D field theory  
$ tr_{m,n} ( P_{ R \bar S } \Phi \otimes \Phi^* ) $ as well 
as related operators in the reduced Matrix quantum mechanics. 
The notation $ P^{\gamma } $ will be used for general 
$\gamma$,   $ P_{ R \bar S }$ being reserved for the special case 
of $ \gamma $ with $k =0 $. Note that, since $ \Phi $ is a general 
complex matrix, not necessarily  satisfying $ \Phi \Phi^{\dagger} =1 $, 
 $ tr_{m,n} ( P_{ R \bar S } \Phi \otimes \Phi^* ) $ cannot be obtained 
just by replacing $ U \rightarrow \Phi $ in $ \chi_{R \bar S } ( U ) $. 

In this section and the next, we will be describing various formulae 
for the construction of the orthogonal set of central projectors 
 $P^{\gamma} $. For $ k \ne 0 $, 
the $P^{\gamma} $  will be a sum of orthogonal {\it symmetric projectors}. 
 \bea 
P^{ \gamma } = \sum_{ A , i } P^ {\gamma }_{ A , i } 
\eea 
The symmetric projectors $P^{\gamma }_{ A , i }$  are not, in general,
 Brauer central elements, 
but they do commute with elements $h$ in the $ \mathbb C [ S_m \times S_n ]$
subalgebra of $B_N (m , n )$. $ A $ labels irreps of 
  $ \mathbb C [ S_m \times S_n ]$.


\subsection{Projector for $k=0$ using character formula}

Starting from (\ref{gencentproj})
we  rewrite the projector for $k=0$ using 
the character formula for elements of Brauer algebra, 
which is in theorem $7.20$ in \cite{halverson}
\begin{equation}\label{char} 
\chi_{B_N {(m,n)}}^{\gamma}(\zeta)
=
N^{h}\sum_{\lambda\vdash m^{\prime}, \pi\vdash n^{\prime}}
\left(
\sum_{\delta \vdash (k-h)} 
g ( \delta,  \gamma^{+} ;   \lambda )
g ( \delta  , \gamma^{-}  ; \pi ) 
\right)
\chi_{S_{m^{\prime}}}^{\lambda }(\zeta^{+})
\chi_{S_{n^{\prime}}}^{\pi} (\zeta^{-})
\end{equation}
where $\gamma^{+}\vdash (m-k)$, $\gamma^{-}\vdash (n-k)$, 
$\zeta^{+}\vdash m^{\prime} =(m-h)$ and 
$\zeta^{-}\vdash n^{\prime} =(n-h)$. 
$\zeta$ denotes a element of Brauer algebra, and we use 
$b$ hereafter instead of $\zeta$. 
$h$ is an integer which is determined by the minimal number of contractions
in a Brauer element. 
For  $b =\sigma\otimes \tau\in 
\mathbb { C }[S_{m}\times S_{n}]$ we have  $h=0$
and $h\ge 1$ if $b$  contains contractions. 

We set $k=0$ in the character formula (\ref{char}).  In this case, 
$\gamma^{+}\equiv R\vdash m$, $\gamma^{-}\equiv S\vdash n$.  
For $b$ with $h\ge 1$, $k-h=-h<0$, so it cannot have 
any partitions $\delta$, hence 
relevant characters vanish. 
Therefore $\chi_{\gamma}(b)\neq 0$ only for
$b\in S_{m}\times S_{n}$ when $k=0$. 
Then the character of $b =\sigma\otimes \tau\in S_{m}\times S_{n}$ 
can be calculated as 
\begin{eqnarray}
\chi_{B_N(m,n)}^{\gamma}(\sigma\otimes\tau)
&=&
\sum_{\lambda\vdash m, \pi\vdash n}
 g  ( \emptyset ,  R ; \lambda ) 
 g ( \emptyset , S ; \pi ) 
\chi_{S_{m}}^{\lambda}(\sigma)
\chi_{S_{n}}^{\pi}(\tau) \cr
&=&
\chi_{S_{m}}^{R}(\sigma)
\chi_{S_{n}}^{S}(\tau) 
\end{eqnarray}

We also use 
\begin{equation}
b^{\ast}=(1^{\ast})b^{-1}  \quad b\in \mathbb { C }[S_{m}\times S_{n}]
\end{equation}
which is a special case of (\ref{bstar}). 
We now rewrite the projector for $k =0$ using the above things. 
\begin{eqnarray}\label{abvthngs} 
P_{R\bar{S}}
&=&
Dim R \bar S \sum_{b}\chi_{\gamma}(b)b^{\ast}\cr
&=&
 Dim R \bar S  \sum_{b\in S_{m}\times S_{n}}
\chi_{\gamma}(b)b^{\ast}\cr
&=&
Dim R \bar S \sum_{\sigma\in S_{m},\tau\in S_{n}}
\chi_{R}(\sigma)
\chi_{S}(\tau) 
1^{\ast} (\sigma\otimes\tau)^{-1}\cr 
&=&
Dim R \bar S
1^{\ast}
\sum_{\sigma\in S_{m}}
\chi_{R}(\sigma)\sigma^{-1}
\sum_{\tau\in S_{n}}
\chi_{S}(\tau) 
\tau^{-1}\cr 
&=&
Dim R \bar S
\frac{m!n!}{d_{R}d_{S}}
1^{\ast}
p_{R}\bar{p}_{S}
\label{rewrittenprojector}
\end{eqnarray}
Using 
\begin{eqnarray}
1^{\ast}&=&
\frac{1}{N^{m+n}}\Sigma^{-1}
\left(\Omega_{m+n}^{-1}\right) \cr
&=&\frac{1}{(m+n)!}
\sum_{ T }\frac{d_{T }}{dim T }
\Sigma^{-1}
\left(p_{ T }\right) \cr
&=&
\frac{1}{\left((m+n)!\right)^{2}}
\sum_{ T \vdash m+n }\frac{d_{T}^{2}}{dim T }
\sum_{\sigma}\chi_{T }(\sigma^{-1})
\Sigma^{-1}
\left( \sigma \right)
\label{1ast}
\end{eqnarray}
where we have used (\ref{omeginv}) to obtain the second line, 
we have 
\bea\label{reproj}  
P_{ R \bar S } 
= \frac{ m!n!}{\left((m+n)!\right)^{2}}
\frac{Dim R \bar S}{d_{R}d_{S}}
\sum_{ T \vdash m+n }\frac{d_{T}^{2}}{dim T }
\sum_{\sigma}\chi_{T }(\sigma^{-1})
\Sigma^{-1}
\left( \sigma \right)
p_{R}\bar{p}_{S}
\eea 

In appendix \ref{sec:calculationof1ast}, we will use  
the expression (\ref{reproj}) to obtain some examples of 
projectors. 
If we take a trace of the expression for $P_{R \bar S } $ 
 we obtain 
\bea 
{ ( m + n )! \over m! n ! } d_R d_S = \sum_{ T } g ( R , S ; T ) d_T 
\eea 
which we know to be a true identity from facts about induced 
representations from $ S_m \times S_n $ to $ S_{ m + n } $
\cite{fulhar}. 
This gives a check of the validity of 
$ t_{\gamma } = Dim  R \bar S  $.

\subsection{ Relation between $B_N(m,n)$ and 
$ \mathbb C [S_{m+n}]$ and a new formula 
for dimension of coupled representations }

Starting with the form 
\bea 
 P _{ \gamma } =t_{\gamma} \sum_{ b  } \chi_{\gamma } ( b^* ) b 
\nnm 
\eea 
we can write the projector as 
\bea 
P_{ \gamma  } = 
 (  Dim \gamma ) \sum_{ \sigma } \Sigma^{-1} 
 ( \sigma ) N^{-m-n } 
           \chi_{ \gamma  } ( \Sigma^{-1} ( \Omega_{m+n }^{-1}\cdot\sigma^{-1}  ) 
  ) 
\eea 
For the $ k=0  $ representations, 
\bea\label{pchibsta}  
P_{ R \bar S   } = (  Dim R \bar S  ) \sum_{ \sigma } N^{-m-n } 
           \chi_{ R \otimes S  }
 ( \Sigma^{-1} ( \Omega_{ m+n }^{-1}\cdot\sigma^{-1}) |_{S_m \times S_n }  ) ~~ 
  \Sigma^{-1} 
 ( \sigma )
\eea
We know that the term without contractions is 
$p_R \otimes p_S $. The coefficient of $1$ is 
$ { d_R^2 d_S^2 \over m! n! } $.  By equating this to 
the term obtained from  (\ref{pchibsta})  by setting $\sigma =1 $, 
we have 
\bea\label{coeffid} 
{ d_R^2 d_S^2 \over m! n! } = N^{-m -n } ~ Dim R \bar S ~ 
\chi_{ R \otimes S } (  \Sigma^{-1} ( \Omega_{m+n}^{-1} )
 \vert_{S_m \times S_n }  ) 
\eea  
Using (\ref{omeginv}) and restricting to the subgroup 
\bea  
\Omega_{m+n}^{-1}\vert_{S_m \times S_n }  = 
 {  N^{m+n} \over  (( m+n )!)^2 }  \sum_T 
 { d_T^2 \over Dim T } \sum_{ \sigma_1 \in S_{m} }  \sum_{ \sigma_2 \in S_{n} }
 \chi_T ( \sigma_1 \cdot \sigma_2  ) \sigma_1 \cdot \sigma_2 
\eea 
The character    $ \chi_T ( \sigma_1 \cdot \sigma_2  )$ can be expanded 
using LR coefficients : 
\bea 
\chi_T ( \sigma_1 \cdot \sigma_2  ) 
= \sum_{ R_1 , S_1  } g ( R_1, S_1 ; T ) \chi_{R_1} ( \sigma_1 ) 
 \chi_{S_1} ( \sigma_2 )  
\eea 
We also need to use
\bea 
&&  \Sigma^{-1} ( \sigma_1 \cdot \sigma_2 ) =  \sigma_1 \cdot
 \sigma_2^{-1}  \cr 
&&  \chi_{ R \otimes S } ( \sigma_1 \cdot \sigma_2^{-1} ) 
   = \chi_R ( \sigma_1 ) \chi_S ( \sigma_2 ) 
\eea  
and the orthogonality of characters 
\bea 
{ 1 \over m!  }
 \sum_{ \sigma_1 } \chi_R ( \sigma_1 ) \chi_{R_1} ( \sigma_1 ) 
= \delta_{ R R_1 } 
\eea 
Using these facts to simplify the RHS of (\ref{coeffid}) 
\begin{equation}\label{newcoupdimform}  
\fbox{
$\displaystyle{
 { d_R^2  d_S^2 
\over  { Dim R \bar S } } =    { m!^2 n!^2 \over { ( m+n )!^2 }} 
\sum_{ T }    { d_T^2  \over Dim T }  g(R, S ; T ) 
}$
}
\end{equation}
For $ m + n \le N $, $T$ above runs over all Young diagrams with 
$m+n$ boxes. In the case $ m +n > N $, but with the condition 
$ c_1 ( R) + c_1 ( S ) < N $ which is necessary for 
$P_{R \bar S } $ to exist, the formula (\ref{newcoupdimform})
is still valid but $T$ now runs over all Young diagrams 
with no more than $N$ rows, or equivalently $ c_1 ( T ) \le N $. 
 We can check (\ref{newcoupdimform})  easily 
in cases such as 
$  ( R, S )  = ( [1], [1] ) ,  ( [ 1] , [2] ) , ( [1] , [2,1] ) , 
( [2] , [ 2,1 ] )$.

\section{ Examples of projectors  } 
\label{examplesprojector}
In this section, we give some examples of projectors. 
The derivation of $k=0$ projectors for some cases based on (\ref{reproj}) 
is given in appendix \ref{sec:calculationof1ast}. 

\subsection{  $ V^{\otimes m } \otimes \bar V $ }\label{vmv1}  

We will now give the general $k=0$ central projector 
for $ B_{N} ( m , 1 ) $ 
\bea\label{pbm1}  
P_{R\bar{[1]}}
=
\left(
1-\frac{1}{N\Omega_{m}}
\sum_{i}\Omega_{m-1}^{<i>}C_{i\bar{1}}
\right)
p_{R}
\eea 
$\Omega_{m-1}^{<i>}$ is the omega factor for 
the $i$-th embedding of $S_{m-1}$ in $S_{m}$, 
where the $i$-th index is removed from $S_{m}$. 
$\Omega_{m-1}^{<i>}$ satisfies 
\begin{eqnarray}
\Omega_{m-1}^{<i>}C_{i\bar{1}}
=
C_{i\bar{1}}\Omega_{m-1}^{<i>}
\end{eqnarray}
and 
\begin{eqnarray}
(ki)\Omega_{m-1}^{<i>}=\Omega_{m-1}^{<k>}(ki) 
\end{eqnarray}
If we use  
\bea
\Omega_{m}=
\Omega_{m-1}^{<k>}
\left(
1+\frac{1}{N}\sum_{i\neq k}^{m}(ik)
\right) 
\label{omegamm-1}
\eea
which was found useful in the study of loop equations 
in 2dYM \cite{srwil},  we get another expression of the projector 
\bea
P_{R\bar{[1]}}
=
\left(
1-\frac{1}{N+\sum_{i\neq k}^{m}(ik)}C_{i\bar{1}}
\right)
p_{R} 
\eea 

We show 
that the projector satisfies $C_{k\bar{1}}P_{R\bar{[1]}}=0$. 
First, we obtain the following equation
\begin{eqnarray}
C_{k\bar{1}}\sum_{i}\Omega_{m-1}^{<i>}C_{i\bar{1}}
&=&
C_{k\bar{1}}\left(\Omega_{m-1}^{<k>}C_{k\bar{1}}
+\sum_{i\neq k}\Omega_{m-1}^{<i>}C_{i\bar{1}}
\right) \cr
&=&
\Omega_{m-1}^{<k>}NC_{k\bar{1}}
+\sum_{i\neq k}C_{k\bar{1}}(ki)\Omega_{m-1}^{<i>} \cr
&=&
C_{k\bar{1}}\Omega_{m-1}^{<k>}N
\left(1+\frac{1}{N}\sum_{i\neq k}(ki)
\right) \cr
&=&
C_{k\bar{1}}\Omega_{m}N 
\end{eqnarray}
where we have used 
(\ref{omegamm-1}). 
Using this equation, it is easy to show 
\begin{eqnarray}
C_{k\bar{1}}P_{R\bar{[1]}}
&=&
\left(
C_{k\bar{1}}-\frac{1}{N\Omega_{m}}
C_{k\bar{1}}\sum_{i}\Omega_{m-1}^{<i>}C_{i\bar{1}}
\right)
p_{R} \cr
&=& 
\left(
C_{k\bar{1}}-\frac{1}{N\Omega_{m}}
C_{k\bar{1}}\Omega_{m}N
\right)
p_{R} \cr
&=& 0 
\end{eqnarray}

We will now show, using (\ref{tpdimf}), that this agrees with the 
Gross-Taylor dimension formula \cite{grotay2}. The trace of 
the first term in (\ref{pbm1}) is  
 \begin{eqnarray}
tr_{m,1}(p_{R})
&=&\frac{d_{R}}{m!}\sum_{R}\chi_{R}(\sigma)
tr_{m,1}(\sigma) \cr
&=&\frac{d_{R}}{m!}\sum_{R}\chi_{R}(\sigma)
N^{K_{\sigma}+1}
\end{eqnarray}
The trace of the second term in (\ref{pbm1}) is 
\begin{eqnarray}
tr_{m,1}\left(
\frac{1}{N\Omega_{m}}\Omega_{m-1}^{<i>}C_{i\bar{1}}
p_{R}\right)
&=&
\frac{d_{R}}{N\chi_{R}(\Omega_{m})}
\sum_{i,j}
tr_{m,1}
(\Omega_{m-1}^{<i>}C_{i\bar{1}}p_{R}) \cr
&=&
\frac{d_{R}}{N\chi_{R}(\Omega_{m})}
\sum_{i}
tr_{m}(\Omega_{m-1}^{<i>}p_{R}) \cr
&=&
\frac{d_{R}}{N\chi_{R}(\Omega_{m})}
DimR
\sum_{i}
\chi_{R}(\Omega_{m-1}^{<i>}) \cr
&=&
\frac{d_{R}}{m!}
N^{m-1}
\sum_{i}
\chi_{R}(\Omega_{m-1}^{<i>}) \cr
&=&
\frac{d_{R}}{m!}
\sum_{\sigma}
\chi_{R}(\sigma)N^{K_{\sigma}+1}
\sigma_{1}\frac{1}{N^{2}} 
\end{eqnarray}
In the last step, we have used the following equation, 
\begin{eqnarray}
\sum_{i=1}^{m}\chi_{R}(\Omega_{m-1}^{<i>}) 
&=&
\sum_{i=1}^{m}\sum_{\sigma\in S_{m-1}^{<i>}}\chi(\sigma)
N^{K_{\sigma}-m}
\cr
&=&\sum_{\sigma\in S_{m}}\sigma_{1}\chi(\sigma)N^{K_{\sigma}-m}.
\end{eqnarray}
where $ \sigma_1 $ is the number of $1$-cycles in $ \sigma$. 
Hence 
\begin{eqnarray}
tr_{m,1}(P_{R\bar{[1]}})
=
\frac{d_{R}}{m!}\sum_{R}\chi_{R}(\sigma)
N^{K_{\sigma}+1}
\left(1-\frac{\sigma_{1}}{N^{2}}\right)
\end{eqnarray}
which can be recognised as $ d_R Dim R [\bar 1 ] $ 
using \cite{grotay2}.

\subsection{ Specific Examples : 
$ V^{\otimes 2 } \otimes \bar V $ }
\label{v2v1}
In this case, we have two $k=0$ projectors 
\begin{eqnarray}\label{b21k0} 
&&P_{[2]\bar{[1]}}
=
\left(1-\frac{1}{N+1}C\right)p_{[2]}  \cr
&&P_{[1^{2}]\bar{[1]}}
=
\left(1-\frac{1}{N-1}C\right)p_{[1^{2}]} 
\end{eqnarray}
where $C\equiv C_{1\bar{1}}+C_{2\bar{1}}$ commutes with any element in 
$ \mathbb C [ S_2]$. 
A $k=1$ projector is given by the sum of the second terms of 
$k=0$ projectors 
\begin{eqnarray}
P^{ ( k=1 , \gamma_{+}=[1],\gamma_{-}=\emptyset )}
= \frac{1}{N+s}C 
= \sum_{R}P_{R\bar{[1]}}^{(k=1,\gamma_{+}=[1],\gamma_{-}=\emptyset)}
\end{eqnarray}
where 
\begin{eqnarray}
&&
P_{[2]\bar{[1]}}^{ ( k=1 , \gamma_{+}=[1],\gamma_{-}=\emptyset)}
=\frac{1}{N+1}Cp_{[2]} \cr
&&
P_{[1^{2}]\bar{[1]}}^{(k=1 , \gamma_{+}=[1],\gamma_{-}=\emptyset) }
=\frac{1}{N-1}Cp_{[1^{2}]} 
\end{eqnarray}
which are symmetric projectors. 

\subsection{ Specific Examples : $ V^{\otimes 3 } \otimes \bar V $ }
\label{b31proj} 
In this case, we have three $k=0$ projectors 
\begin{eqnarray}
&&P_{[3]\bar{[1]}}
=
\left(1-\frac{1}{N+2}C\right)p_{[3]}  \cr
&&P_{[1^{3}]\bar{[1]}}
=
\left(1-\frac{1}{N-2}C\right)p_{[1^{3}]}  \cr
\cr
&&P_{[2,1]\bar{[1]}}
=\left(1
-\frac{N}{(N^{2}-1)}C-\frac{1}{(N^{2}-1)}D\right)p_{[2,1]}
\label{t=0m3n1}
\end{eqnarray}
where 
$C\equiv C_{1\bar{1}}+C_{2\bar{1}}+C_{3\bar{1}}$ and 
$D\equiv C_{1\bar{1}}s_{2}+C_{2\bar{1}}s_{1}s_{2}s_{1}+C_{3\bar{1}}s_{1}$ 
commute with any element in $ \mathbb C [ S_3]$.
Some useful formulae are given in appendix \ref{sec:appB(31)}

We have two $k=1$ projectors 
\begin{eqnarray}\label{centraldecomp1} 
P^{ ( k=1 ,\gamma_{+} =[2] , \gamma_{-} =  \emptyset   )}
&=&
\frac{1}{N+2}Cp_{[3]}
+\frac{1}{2}\frac{1}{N-1}(C+D)p_{[2,1]}\cr
{\rule[-2mm]{0mm}{8mm}\ } 
&=&
P_{[3]\bar{[1]}}^{ ( k=1 ,\gamma_{+} =[2] , \gamma_{-} =  \emptyset   )}
+P_{[2,1]\bar{[1]}}^{ ( k=1 ,\gamma_{+} =[2] , \gamma_{-} =  \emptyset   )}
\cr
{\rule[-2mm]{0mm}{9mm}\ } 
P^{ ( k=1 , \gamma_{+} = [1^{2}], \gamma_{-} =  \emptyset)     }
&=&
\frac{1}{N-2}C p_{[1^{3}]}
+\frac{1}{2}\frac{1}{N+1}(C-D)p_{[2,1]}\cr
{\rule[-2mm]{0mm}{8mm}\ } 
&=&
P_{[1^{3}]\bar{[1]}}
^{ ( k=1 ,\gamma_{+} =[1^{2}] , \gamma_{-} =  \emptyset   )}
+P_{[2,1]\bar{[1]}}
^{ ( k=1 ,\gamma_{+} =[1^{2}] , \gamma_{-} =  \emptyset   )}
\end{eqnarray}
Each term in the RHS is a symmetric projector. 
Therefore, we have seven symmetric operators and 
five central Brauer projectors. The decomposition of central Brauer 
projectors into symmetric projectors 
 in (\ref{centraldecomp1}) 
can be understood in terms of the decomposition of Brauer irreps 
in terms $ \mathbb C [ S_3 \times S_1 ]$ irreps. For example, 
we know using (\ref{redcoeff}) that 
the   $( k=1 ,\gamma_{+} =[2] , \gamma_{-} =  \emptyset   )$ irrep 
of $B_N(3,1)$  contains 
the direct sum of  irreps $ ( [3] , [1] ) $ and $ ( [2,1] , [1] )  $  of  
$ \mathbb C [ S_3 \times S_1 ]$, each with unit multiplicity. 

\subsection{ Specific examples : $ V^{\otimes 2 } \otimes \bar V^{\otimes 2 } $  }\label{b22proj} 
We have four $k=0$ projectors 
\begin{eqnarray}
P_{R\bar{S}}
&=&
\left(1
-\frac{1}{(N+s+\bar{s})}C_{(1)} 
+\frac{1}{(N+s)(N+s+\bar{s})}C_{(2)}\right)p_{R}\bar{p}_{S} 
\label{t=0form2n2}
\end{eqnarray}
where 
$R=[2]$ or $[1^{2}]$, $S=[2]$ or $[1^{2}]$, 
$s=(12)$, $\bar{s}=(\bar{1}\bar{2})$ and 
\begin{equation}
C_{(1)}=C_{1\bar{1}}+C_{1\bar{2}}+C_{2\bar{1}}+C_{2\bar{2}} \quad
C_{(2)}=C_{1\bar{1}}C_{2\bar{2}}+C_{1\bar{2}}C_{2\bar{1}} 
\end{equation}
which commute with any element in $\mathbb C [ S_2 \times S_2] $.  
Some useful formulae are given in appendix \ref{sec:projector22}. 
$k\neq 0$ central Brauer projectors are given by 
\begin{eqnarray}
&&
P^{ ( k=1 , \gamma_+ = [1], \gamma_- = [1]  ) }=
\frac{1}{(N+s+\bar{s})}\left(C _{(1)}
-\frac{2}{N}C_{(2)}\right) 
\cr
&&
P^{ (k=2 ,  \gamma_+ = \emptyset, \gamma_- = \emptyset  )   }=
\frac{1}{N(N+s)}C_{(2)}
\end{eqnarray}
These two $k\neq 0$ central Brauer projectors can be written as the sum of 
symmetric projectors as 
\begin{eqnarray}
&&
P^{ ( k=1 , \gamma_+ = [1], \gamma_- = [1] )  }=
\sum_{R,S}P^{ ( k=1 , \gamma_+ = [1], \gamma_- = [1] ) }_{R\bar{S}}
\cr
&&
P^{ (k=2 ,  \gamma_+ = \emptyset, \gamma_- = \emptyset  )   }=
\sum_{R,S}P^{(k=2 ,  \gamma_+ = \emptyset, \gamma_- = \emptyset  )   }_{R\bar{S}}
\end{eqnarray}
where 
\begin{eqnarray}
&&
P^{ ( k=1 , \gamma_+ = [1], \gamma_- = [1] )  }_{R\bar{S}}
=
\frac{1}{(N+s+\bar{s})}\left(C_{(1)} 
-\frac{2}{N}C_{(2)}\right)p_{R}\bar{p}_{S}
\cr
&&
P^{(k=2 ,  \gamma_+ = \emptyset, \gamma_- = \emptyset  ) }_{R\bar{S}}
=
\frac{1}{N(N+s)}C_{(2)}p_{R}\bar{p}_{S}
\end{eqnarray}
Because of 
\begin{eqnarray}
C_{(2)}p_{[2]}\bar{p}_{[1,1]}=C_{(2)}p_{[1,1]}\bar{p}_{[2]}=0 
\end{eqnarray}
which can be easily checked using $C_{(2)}s=C_{(2)}\bar{s}$, 
$P^{ (k=2 ,  \gamma_+ = \emptyset, \gamma_- = \emptyset  )   }
_{[2]\bar{[1^{2}]}}$ and 
$P^{ (k=2 ,  \gamma_+ = \emptyset, \gamma_- = \emptyset  )   }
_{[1^{2}]\bar{[2]}}$ vanish. 
Therefore we have ten symmetric projectors and 
six central Brauer projectors.


\subsection{ Specific examples : Composites of symmetric and anti-symmetric } 
We write down projectors 
corresponding to 
an AdS giant and an AdS anti-giant $ ( [m ],[ n ] ) $,  an  S-giant 
and an S-anti-giant $ ( [1^{m}] ,[ 1^{n}]  ) $,
 and a composite of an AdS giant and an S anti-giant
$ ( [m] , [1^{n} ]) $ or vice versa $ ( [1^{m}] , [n] ) $.
We assume $m\ge n$ in this subsection. 

We first define 
\begin{eqnarray}
&&
C_{(1)}=\sum_{ij}C_{i\bar{j}} \quad
C_{(2)}=\frac{1}{2!}\sum_{i\neq j }\sum_{k\neq l}C_{i\bar{k}}C_{j\bar{l}} 
\quad 
\cdots 
\cr
&&
C_{(k)}=\frac{1}{k!}\sum_{i_{a}\neq i_{b}}\sum_{j_{a}\neq j_{b}}
C_{i_{1}\bar{j_{1}}}C_{i_{2}\bar{j_{2}}}\cdots C_{i_{k}\bar{j_{k}}}
\end{eqnarray}
Using these, we obtain projectors for $k=0$ representations: \\
$(R,S)=([m],[n])$ 
\begin{eqnarray}
P_{[m]\bar{[n]}}
&=&
\left(1-\frac{1}{N+m+n-2}C_{(1)} \right.\cr
&& \left. \qquad
+\frac{1}{(N+m+n-3)(N+m+n-2)}C_{(2)}
+\cdots
\right)
p_{[m]}\bar{p}_{[n]} \cr
&=&
\left(
1+\sum_{k=1}^{n}(-1)^{k}\prod_{l=1}^{k}
\frac{1}{(N+m+n-l-1)}C_{(k)}
\right)
p_{[m]}\bar{p}_{[n]}
\end{eqnarray}
$(R,S)=([1^{m}],[1^{n}])$ 
\begin{eqnarray}
P_{[1^{m}]\bar{[1^{n}]}}
&=&
\left(
1+\sum_{k=1}^{n}(-1)^{k}\prod_{l=1}^{k}
\frac{1}{(N-m-n+l+1)}C_{(k)}
\right)
p_{[1^m]}\bar{p}_{[1^n]}
\end{eqnarray}
$(R,S)=([m],[1^{n}])$ 
\begin{eqnarray}
P_{[m]\bar{[1^{n}]}}
&=&
\left(
1+\sum_{k=1}^{n}(-1)^{k}\prod_{l=1}^{k}
\frac{1}{(N+m-n)}C_{(k)}
\right)
p_{[m]}\bar{p}_{[1^n]}
\end{eqnarray}
$(R,S)=([1^{m}],[n])$ 
\begin{eqnarray}
P_{[1^{m}]\bar{[n]}}
&=&
\left(
1+\sum_{k=1}^{n}(-1)^{k}\prod_{l=1}^{k}
\frac{1}{(N-m+n)}C_{(k)}
\right)
p_{[1^m]}\bar{p}_{[n]}
\end{eqnarray}

In appendix 
\ref{sec:CP=0forsymandantisym}, we give a proof of 
$C_{i\bar{j}}P_{R\bar{S}}=0$ for these projectors. 
These expressions correctly reduce to the relevant examples 
from subsections \ref{vmv1}-\ref{b22proj}

\section{ Counting of operators  } 
\label{sec:countingoperator}

As observed in section 3.4 Brauer elements can be used to construct 
multi-trace local operators from complex matrices $ \Phi , \Phi^{\dagger} $. 
 We also observed that the counting of 
these multi-trace operators is the same as counting symmetric elements 
in $B_N(m,n)$, which are elements that commute with the
$  \mathbb C [ S_m \times S_n  ] $ sub-algebra of $ B_{N} ( m, n )$. 
A class of symmetric elements are symmetric projectors. 
These projectors have appeared in sections 4 and 5 as 
summands in central Brauer projectors. This relation between 
 central Brauer projectors and symmetric projectors 
corresponds to the group theory counting
of irreps of $ B_{N} ( m, n ) $ in $ \vmn $ weighted 
by the multiplicity of  $  \mathbb C [ S_m \times S_n  ] $
irreps. So the number of symmetric projectors $N_s ( m,n) $ is given 
by 
\bea 
N_s ( m,n ) = \sum_{ \gamma } \sum_{ A } M^{\gamma }_{ A } 
\eea 
where $A $ labels irreps of the symmetric group $ S_m \times S_n $. 
It is given by a pair $  ( \alpha , \beta ) $ which is a pair of 
partitions of $m ,  n $ respectively. Using the expression 
(\ref{redcoeff}) for the multiplicities we obtain 
 \bea\label{countmult}  
&& N_{s}  ( m , n ) = 
\sum_{k=0}^{min(m,n)} \sum_{ \gamma_+ \vdash (m-k) }
 \sum_{ \gamma_- \vdash (n-k) } \sum_{ \alpha \vdash m } 
\sum_{  \beta \vdash n  }     \left( ~ \sum_{ \delta \vdash k }
  g ( \delta , \gamma_+ ; \alpha  ) 
g ( \delta , \gamma_- ; \beta ) ~  \right)   
\eea 
The most general symmetric element of $B_N(m,n)$ is not 
necessarily a projector, but we can argue that the 
most general element is a symmetric branching operator, 
a special case of which is a projector. These will be discussed 
in more detail in section 7. They are counted by a  $N_{sb} $ 
\bea\label{nsb}  
 N_{ sb } &=& \sum_{\gamma } \sum_{ A }  ( M^{ \gamma  }_ { A } )^2  \cr 
&=& \sum_{k=0}^{min(m,n)} \sum_{ \gamma_+ \vdash (m-k) }
 \sum_{ \gamma_- \vdash (n-k) } \sum_{ \alpha \vdash m } 
\sum_{  \beta \vdash n  }     \left( \sum_{ \delta \vdash k }
  g ( \delta , \gamma_+ ; \alpha  ) 
g ( \delta , \gamma_- ; \beta )  \right)^2   
\eea

  At large $N$, i.e $ m + n < N $ 
the counting of traces is equivalent to a problem of 
counting necklaces with coloured beads, which is solved by Polya theory. 
In this large $N$ case, the counting of gauge invariant 
operators by Polya theory agrees with the counting in terms of 
Brauer algebras. 
When we drop the restriction and deal with  
finite $N$ effects, the connection to Brauer algebras allows a simple solution 
of the finite $N$ counting problem. The sums over Brauer irreps 
are reduced to those which appear in the decomposition of 
$ V^{  \otimes m } \otimes \bar V^{ \otimes n } $.
 This will be discussed further in section 8.
For other recent discussions of finite $N$ matrix counting problems 
see \cite{fhh,dol,ammpv}.

The counting of traces ( at large $N$ ) is given
by Polya theory as 
\bea\label{polya}  
T ( x , y ) & =& \prod_{n=1}^{\infty } { 1 \over ( 1-   x^n - y^n )  }  \cr 
            &=& \sum_{ m =0 , n =0}^{  \infty }  t ( m , n ) x^m y^n 
\eea 
The coefficients $ t ( m,n ) $ count the number of traces 
with $ m $ copies of $ \Phi$ and $ n $ copies of $ \Phi^{  \dagger } $. 
For a recent discussion of this in the physics literature see 
\cite{pauletal}.  We are led,  from the above discussion, to  
\bea\label{bcount}  
N_{sb} ( m,n  ) = t ( m ,n )
\eea

There are a number of interesting cases where the multiplicities
$ M^{\gamma}_{A }=  M^{\gamma}_{ \alpha , \beta  }$ are all either 
$1$ or $0$. In these cases $ N_s ( m, n ) = N_{ sb } ( m ,n )  = t ( m ,n )$. 
One class of such examples is  $ B_N( m ,1 ) $. In this case  
 $ t(m,1) $ can be obtained by calculating the 
derivative of  $ { \partial T ( x, y ) \over \partial y } |_{ y=0}  $.
\bea 
&&  { \partial T ( x, y ) \over \partial y } |_{ y=0}  \cr
&&  = { 1 \over 1 - x }  \prod_{ n=1}^{\infty } { 1 \over ( 1 - x^n ) }  \cr 
&&  = \sum_{ m_1=0}^{\infty }  x^{m_1} \sum_{ m_2 =0}^{\infty} p ( m_ 2)
 x^{ m_2 }    \cr 
&& = \sum_{ m =0 }^{\infty } x^m \sum_{ k=0 }^m p ( k  )     
\eea 
where $ p(k)$ is the number of partitions of $k$. Hence 
\bea 
t ( m , 1 ) = \sum_{ k=0 }^{ m } p ( k  ) 
\eea 
This satisfies a recursion relation
\bea 
t ( m +1 , 1 ) = p ( m +1 ) + t ( m, 1 )
\eea 
The same recursion relation can be derived for 
$ N_s ( m,1 ) $. The sum over $k$ for 
$ N_s  ( m, 1) $ has two terms. The $k=0$ term gives 
$ p ( m ) p(1) = p(m )$. To get the second term, we 
sum over Young diagrams  $ R  $ of $ m $ boxes.
Let such a diagram have $c_1 $ columns of length 
$1$, $c_2$ columns of length $2$ etc. In other 
words $j$ corresponds to column length and $c_j$ 
gives to multiplicity of that column length in the 
Young diagram of $R$.   
So $ m = \sum_{j=0}^m j c_j $. For each Young diagram 
the factor $ \sum_R \sum_{ \gamma_+ \vdash (m-1) }
 g ( \gamma_+  , [1] ; R )  $ 
is equal to the number of ways of removing a box 
from $R$ to get a legal Young diagram of $m-1$ boxes. 
This can be  seen to be equal to the number of non-zero 
column length multiplicities $c_j$'s. 
Hence, 
\bea
 N_s ( m,1 ) = \sum_{ R \vdash m }  ( 1 +
\hbox{  number of ways of removing a box from $R$ }  ) 
\eea 
Now $ ( 1 + \hbox { number of ways of removing a box from $R$ } ) $ is 
equal to number of way of adding a box to $R$. Hence 
\bea 
 N_s ( m ,1 )
& = &\sum_{ R \vdash m } ( 1 + 
\hbox { number of ways of removing a box from $R$ }  ) \cr 
& =& \sum_{ R \vdash m+1  } (  \hbox { number of ways of removing
 a box from $R$  } )
\eea 
But we also know that  
\bea 
 N_s ( m+1 , 1 ) = p(m+1 ) + 
\sum_{ R \vdash m+1  } 
(  \hbox { number of ways of removing a box from $R$ }  )
\eea 
Hence $ N_s ( m+1 , 1 ) = p ( m +1 ) + N_s ( m ,1 ) $. This the same 
recursion relation as for $  t ( m,1 ) $. It is also 
easily checked that $ N_s ( 1 ,1 ) = t ( 1,1 )$. This proves the desired 
identity between the number of traces and the number of 
Brauer irreps weighted with symmetric group decomposition 
multiplicities.   

We have also checked for $ B_{N} ( m, 2 ) $ in the cases 
$ m = 1\cdots  5 $ that $ N_s ( m , 2 ) = N_{sb} ( m , 2 ) = t ( m , 2 ) $. 
For cases such as  $ B_{N} ( 3, 3 ) $, we find  
 $ t ( 3 , 3  )  = N_{ sb } ( 3, 3)  =  38   $  whereas $ N_s ( 3,3 ) = 36 $. 
We have also checked $ N_{ sb } ( 4,3 ) = t  ( 4,3 )  ; 
 N_{ sb} ( 4,4 ) = t ( 4,4 ) ; 
 N_{ sb } ( 4,5 ) = t  ( 4,5 )  ;  N_{ sb } ( 5,5 ) = t  ( 5,5 )$. 
 Based on these non-trivial examples, and the discussion of
sections \ref{specgen},  \ref{sec:orthogonalsetoperator} 
we expect that (\ref{bcount}) is true in general. At finite 
$N$ there is a cutoff on $\gamma $ in (\ref{nsb}) following from 
(\ref{nonchirschwe}) of $ c_1 ( \gamma_+ ) + c_1 ( \gamma_- ) \le N $.

\section{ Orthogonal set of operators for brane-anti-brane systems.  }
\label{sec:orthogonalsetoperator}
A representation $ \gamma $ of $ B_{N} ( m , n )$ can be  decomposed 
into irreducible representations $ A$ of the 
$ \mathbb C [ S_m \times S_n ] $ sub-algebra. The index $ A$ consists of a
 pair $ ( \alpha , \beta ) $ where $ \alpha $ is a partition of 
$ m $, and $ \beta $ is a partition of 
$n$, which is   expressed as $ \alpha \vdash m  ,  \beta \vdash n $. 
An irrep $A$ will generically appear with multiplicity 
$M^{\gamma}_{A } $, and we will use an index $i$ which 
runs over this multiplicity. Let us denote by
 $ | \gamma ; A , m_A ; i   \rangle  $ an orthonormal set of vectors  
 in the $ \gamma $ representation which transforms 
in the $i $ th copy of the  state $m_A $ of the irrep $A$ of the
 sub-algebra.  
Central projectors for $ \gamma $ in the regular representation 
can be written as 
\bea 
P^{\gamma  } = \sum_{ i } \sum_{A , m_A } | \gamma ; A , m_A ; i \rangle 
 \langle  \gamma ; A , m_A ; i | 
\eea 
The construction of these projectors in terms of the 
algebra has been discussed at length in section 4. 
Define 
\bea 
 P^{\gamma}_{A , i }  = \sum_{ m_A } | \gamma ; A , m_A ; i \rangle 
          \langle \gamma ; A , m_A ; i | 
\eea 
Here we are not summing over $A , i $. 
As we will show these commute with $  \mathbb C [ S_m \times S_n ] $, 
but not in general with $B_N(m,n)$.  
Examples of these symmetric projectors have also been computed 
in section 5. 
These projectors belong to a more general class of symmetric elements. 
\bea 
Q^{\gamma }_{A , i  j } 
= \sum_{ m_A } | \gamma ; A , m_A ; i \rangle 
\langle \gamma ; A , m_A ; j | 
\eea   
Consider 
\bea 
 h Q^{\gamma }_{A , i  j }  h^{-1} 
& =& \sum_{m_A ,n_A , k_A } D^{ A }_{n_A m_A } ( h ) 
| \gamma ; A , n_A ; i \rangle \langle \gamma ; A , k_A ; j | 
 D^{ A }_{m_A k_A } ( h^{-1}  ) \cr 
& =&  \sum_{n_A , k_A } D^{ A }_{n_A k_A } ( 1 ) | \gamma ; A , n_A ; i \rangle \langle \gamma ; A , k_A ; j | \cr 
& =&  \sum_{n_A , k_A } \delta_{ n_A k_A } | \gamma ; A , n_A ; i \rangle \langle \gamma ; A , k_A ; j | \cr 
& =& ~  Q^{\gamma }_{A , i  j }
\eea 
The $D^{ A }_{n_A m_A } ( h )$ are matrix elements of $h$ 
in the irrep $A$. 
This shows that  $Q^{\gamma }_{A , i  j }$ commutes with 
the subalgebra. This property is 
denoted by saying $Q^{\gamma }_{A , i  j }$ are {\bf symmetric 
branching operators}.  By using an expansion of a general element of 
the Brauer algebra 
in terms of matrix elements of irreps as in \cite{ramthesis} we expect 
it should be possible to prove that  the symmetric branching operators
provide a complete set of symmetric elements in the Brauer algebra
This is supported by the counting examples we have done 
in section \ref{sec:countingoperator}. 
 Since $  P^{\gamma}_{A , i } =  Q^{\gamma }_{A , i  i  }$
we also have 
\bea 
h  P^{\gamma}_{A , i } h^{-1} =  P^{\gamma}_{A , i }
\eea 
This property is 
expressed  by saying  $ P^{\gamma}_{A , i } $ are symmetric projectors. 
Using the orthonormality of the states  we can derive 
\bea\label{branchmatmult}  
Q^{\gamma_1}_{A , i  j } Q^{\gamma_2}_{B , k  l }
= \delta_{\gamma_1 \gamma_2} \delta_{A B } \delta_{j k }
 Q^{\gamma_1}_{A , i  l  }
\eea

Associated with the symmetric elements $Q$, we can find a  
 complete basis in the space of local operators 
constructed from $ \Phi , \Phi^{\dagger} $ in four dimensional 
$N=4$ SYM. It also gives a complete basis of gauge invariant
operators built from the matrices $A^{\dagger} , B^{\dagger}$.  
We will show that this basis for operators diagonalises 
the correlators. We first discuss this in the context of the matrix 
quantum mechanics.

\subsection{  Reduced 1D Matrix Model : Orthogonal basis using Brauer } 

It has been shown \cite{cjr, djr,rodrigues} that the reduction of the 
four-dimensional action on  $S^3 \times R $ leads to 
the Hamiltonian and $SO(2) $ symmetry 
\bea 
&& H = tr  ( A^{\dagger } A + B^{\dagger } B )  \cr 
&& J = tr  (  A^{\dagger } A - B^{\dagger } B ) 
\eea 
The matrices obey the algebra 
\bea 
&& [ A^{i}_j , A^{ \dagger  ~ k }_l  ] = \delta^i_l \delta^{k}_j \cr 
&& [  B^{i}_j , B^{ \dagger  ~ k }_l  ] = \delta^i_l \delta^{k}_j \cr 
&& [ A^{i}_j ,  B^{ \dagger  ~ k }_l  ] = [ A^{i}_{j} , B^{k}_l ] 
   = [ A^{\dagger ~ i}_j ,  B^{ \dagger  ~ k }_l  ]
   = [ A^{\dagger ~ i}_j ,  B^{   ~ k }_l  ] = 0 
\eea

Gauge invariant states are obtained by acting 
with traces of $ A^{ \dagger } $ and $ B^{\dagger } $ 
on the vacuum, e.g 
\bea 
&& Tr  ( A^{\dagger} )^n  | 0 > \quad E = J = n \cr 
&& Tr  (  B^{\dagger} )^n   | 0 >  \quad E = - J = n \cr 
&&  Tr ( A^{\dagger } )^n (  B^{\dagger} )^m | 0 > 
  \quad E = n + m , J = n -m 
\eea 

Among the states obtained by acting with 
$ A^{\dagger } $ a complete orthogonal 
set is obtained from the Schur polynomials 
\bea 
\chi_R ( A^{ \dagger}  ) | 0  > 
\eea 
They obey 
\bea 
< 0 | \chi_R ( A ) \chi_S ( A^{ \dagger}  ) | 0  > 
= \delta_{ R S }  { n ! Dim R \over d_R }  
\eea 
We would like to find a complete set of orthogonal states 
in the more general case where both $ A^{\dagger} $ and $ B^{\dagger} $ 
are acting on the vacuum. 

We claim that the operators
 $ tr_{m,n}  \bigl ( \Sigma ( Q^{\gamma }_{A, i j}  \bigr )   
( A^{\dagger }\otimes  B^{\dagger} )  \bigr )| 0 > $ 
 diagonalise the quantum mechanical inner product
\bea\label{brauerdiagcor}  
&& < 0 |  tr_{m,n}   
\bigl ( \Sigma ( Q^{\gamma_2 ~ \dagger }_{A_2 , i_2 j_2} )   
( A\otimes  B )  \bigr )
  tr_{m,n}  \bigl ( \Sigma ( Q^{\gamma_1 }_{A_1 , i_1 j_1} )   
( A^{\dagger }\otimes  B^{\dagger} )  \bigr )| 0 > \cr 
&& = m! n! \delta_{\gamma_1 \gamma_2 } \delta_{A_1 A_2} 
\delta_{i_1 i_2  } \delta_{j_1 j_2 }~ d_{A_1} ~  Dim  \gamma_1 
\eea 

\begin{figure}[t]
\begin{center}
 \resizebox{!}{6cm}{\includegraphics{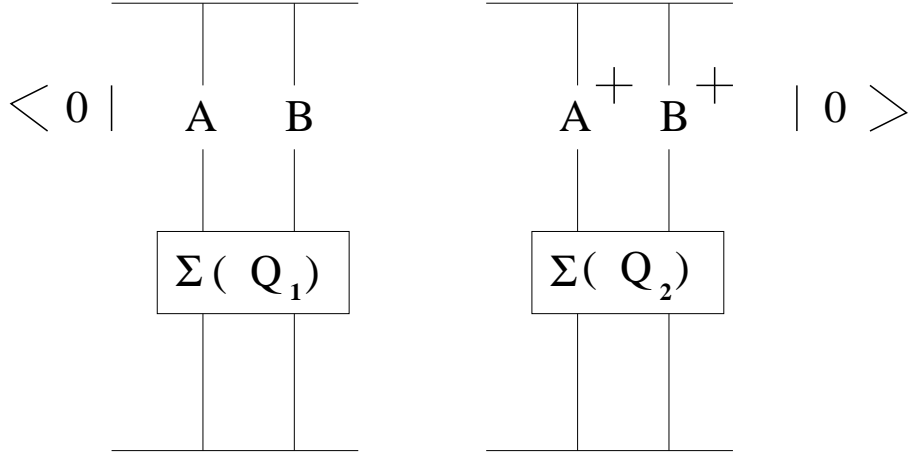}}
\caption{Diagrammatic representation of inner product    }
 \label{fig:innerproductQM}
\end{center}
\end{figure}

\begin{figure}[tbp]
\begin{center}
 \resizebox{!}{6cm}{\includegraphics{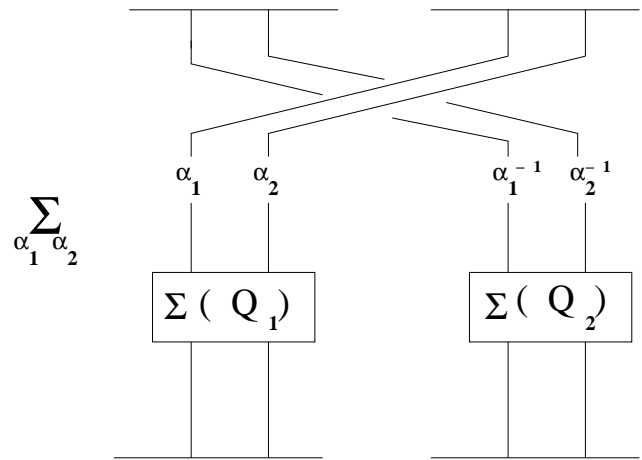}}
\caption{ Sum over all possible contractions leads to a sum
 over permutations in tensor space  }
 \label{fig:innerprodQMperms}
\end{center}
\end{figure}

\begin{figure}
\begin{center}
 \resizebox{!}{8cm}{\includegraphics{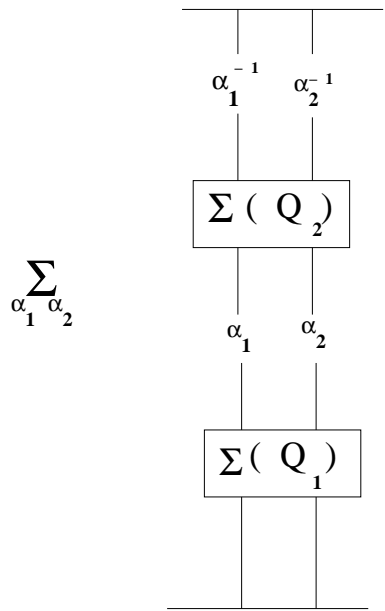}}
\caption{ Tensor space identities allow the diagrammatic simplification  }
 \label{fig:innerprodQMstraighten}
\end{center}
\end{figure}

Consider the LHS of (\ref{brauerdiagcor}). We can describe it diagrammatically 
as in Figure \ref{fig:innerproductQM}. We have used $Q_1 , Q_2 $ 
for $ Q^{\gamma_1 }_{A_1 , i_1 j_1}, 
 Q^{\gamma_2 ~ \dagger }_{A_2 , i_2 j_2} $ in the figures to keep them 
simple. 
It is understood that the upper horizontal  line is 
identified with the lower horizontal line, which expresses the 
identification of tensor space indices for a trace. 
The sum over all Wick contractions gives a sum over permutations 
in tensor space as in Figure \ref{fig:innerprodQMperms}. An  obvious 
diagrammatic manipulation, which corresponds to an identity 
in tensor space, results in  Figure \ref{fig:innerprodQMstraighten}. 
This allows us to write 
\bea\label{pfdiag} 
&& < 0 |  tr_{m,n} \bigl ( \Sigma (
 Q^{\gamma_2  \dagger  }_{A_2 , i_2 j_2} )   
( A \otimes  B )  \bigr )
 tr_{m,n}   \bigl ( \Sigma ( Q^{\gamma_1 }_{A_1 , i_1 j_1} )   
( A^{\dagger }\otimes  B^{\dagger} )  \bigr )| 0 > \cr 
&& = \sum_{ \alpha_1 \in S_{m}} \sum_{ \alpha_2 \in S_n } 
tr_{m, n }   \bigl(  ( \alpha_1 \otimes \alpha_2 )
  \Sigma ( Q^{\gamma_2 ~  \dagger  }_{A_2 , i_2 j_2} )  
( \alpha_1^{-1}  \otimes \alpha_2^{-1} ) 
\Sigma ( Q^{\gamma_1   }_{A_1 , i_1 j_1} ) \bigr ) \cr 
&& =  m! n! tr_{m,n} 
  \bigl( \Sigma ( Q^{\gamma_2  \dagger  }_{A_2 , i_2 j_2} ) 
 \Sigma ( Q^{\gamma_1   }_{A_1 , i_1 j_1 } )  \bigr ) \cr 
&& = m ! n! tr _{m,n} 
\bigl (  Q^{\gamma_2  ~ \dagger   }_{A_2 , i_2  j_2}  
                  Q^{\gamma_1    }_{A_1 , i_1 j_1} \bigr ) \cr 
&& = m! n! \delta_{\gamma_1 \gamma_2 }\delta_{A_1 A_2 } \delta_{i_1 i_2 }  
tr _{m,n} ( Q^{\gamma_1    }_{A_1 , j_2 j_1} ) \cr 
&& = m! n! \delta_{\gamma_1 \gamma_2 }\delta_{A_1 A_2 }
\delta_{i_1 i_2 } \delta_{j_2 j_1 } ~  d_{ A_1 } ~ 
Dim \gamma_1 
\eea 
The second line  follows from the diagrammatics. 
The third  line follows using (\ref{symsym} ). 
The fourth line follows from (\ref{mapbilfor}). 
In the last line we have used the Schur-Weyl duality (\ref{nonchirschwe}). 
The factor $ Dim  \gamma_1  $ is the dimension of the 
$GL(N)$ irrep labelled by $ \gamma_1$ and $d_{A_1} $ is the 
dimension of the corresponding irrep of $\mathbb C [ S_m \times S_n  ] $.
This proves (\ref{brauerdiagcor}).  

Using the relations between the symmetric projectors 
$ P^{\gamma}_{A , i } $ or central projectors $ P^{\gamma} $ 
in terms of these symmetric branching operators, we can derive 
\bea 
< 0  | tr_{m,n}  ( \Sigma ( P^{\gamma_2 }_{A_2 , i_2 } ) ) 
(  A\otimes  B ) tr _{m,n} (   \Sigma ( Q^{\gamma_1 }_{A_1  , i_1 j_1 } )  
 ( A^{\dagger }\otimes  B^{\dagger} ) ) | 0 > 
= m!n! \delta_{\gamma_1 \gamma_2 }  \delta_{ A_1 A_2} 
\delta_{ i_{2}  i_1   } \delta_{i_{2} j_1  }    d_{A_1}  
  Dim \gamma_1
\nnm 
\eea 
This is proportional to  $ \delta_{ i_1 j_1 } $ , which guarantees 
that the $Q^{\gamma_1 }_{A_1  ,  i_1 j_1 } $ overlaps with a symmetric projector if it  is actually itself a symmetric projector. 
Considering the overlap between operators constructed from two symmetric projectors, we have 
\bea 
< 0  | tr_{m,n} ( \Sigma ( P^{\gamma_2 }_{A_2 , i_2 } ) ) 
(  A\otimes  B ) tr_{m,n}  (   \Sigma ( P^{\gamma_1 }_{A_1  ,  i_1 } )  
 ( A^{\dagger }\otimes  B^{\dagger} ) ) | 0 >  \cr
= m!n!
\delta_{\gamma_1 \gamma_2 }  \delta_{ A_1 A_2}  \delta_{i_1 i_2} 
d_{A_1}  
 Dim \gamma_1
\eea 

We can also show that central projectors corresponding to 
different irreps $ \gamma $ give orthogonal states : 
\bea\label{cprojorth} 
&& < 0  | tr_{m,n}  ( \Sigma ( P^{\gamma_2 } ) ) 
(  A\otimes  B ) tr_{m,n}  (   \Sigma ( P^{\gamma_1 } )  
 ( A^{\dagger }\otimes  B^{\dagger} ) ) | 0 >  \cr 
&& ~~~~~~~~ =  m ! n ! \delta_{ \gamma_1 \gamma_2 }  
tr_{m,n}  ( P^{\gamma_1}   )  
= m! n! \delta_{ \gamma_1 \gamma_2 }~  d^{(B)}_{\gamma_1} ~  Dim \gamma_1 
\eea 
In this equation $d^{(B)} $ is a Brauer dimension. 
This can be derived by relating $P^{\gamma} $ to 
the $Q$ operators, or applying the diagrammatics directly 
to the LHS of (\ref{cprojorth}) and using the projector property 
$ P^{\gamma_1} P^{\gamma_2 } = \delta_{\gamma_1 \gamma_2}  P^{\gamma_1} $.

Recall from section 3 that Brauer irrep labels $\gamma$ 
determine an integer $k$ and  partitions 
$ \gamma_+ \vdash m -k , \gamma_- \vdash n-k $.  
For $k=0$, the irrep $\gamma$ decomposes into a unique irrep 
$ \gamma_+ , \gamma_- $ of $ \mathbb C [ S_m \otimes S_n ]$. 
This means that the $k=0$ central projectors do not decompose 
into a sum of multiple symmetric projectors. Another  special 
property of the $k=0$ projectors becomes apparent when we consider the 
field theory operators, rather than Matrix quantum mechanics operators.  

\subsection{ Orthogonal multi-matrix operators in 4D field Theory } 

In the case of 4D field theory we associate gauge invariant 
Matrix operators, much as we do in Matrix quantum mechanics. 
Now we consider $  : tr_{m,n}  
\bigl ( \Sigma ( Q^{\gamma }_{A, i j}  \bigr )   
( \Phi \otimes  \Phi^{\dagger} )  \bigr ) : ~$  . The notation 
$ : \cO : $ for an operator $ \cO $ indicates that we have subtracted the 
short distance singularities, to give an operator which will have
 no self-contractions inside correlators as explained in the 
example in section 2.   The orthogonality 
property of (\ref{brauerdiagcor}) 
has a direct analog, with identical derivation  
\bea\label{brauerdiagcorft}  
&& <   :  tr_{m,n}
\bigl ( \Sigma ( Q^{\gamma_2 ~ \dagger }_{A_2 , i_2 j_2} )   
( \Phi^{\dagger}  \otimes  \Phi  )  \bigr ) : ~
 :   tr _{m,n} \bigl ( \Sigma ( Q^{\gamma_1 }_{A_1 , i_1 j_1} )   
( \Phi \otimes  \Phi^{\dagger} )  \bigr ) :  > \cr 
&& = m! n! \delta_{\gamma_1 \gamma_2 } \delta_{A_1 A_2} 
\delta_{i_1 i_2  } \delta_{j_1 j_2 } ~ d_{A_1 } ~ Dim  \gamma_1    
\eea 
To be more explicit we would add the dependence on $x_1, x_2 $ 
in the two operators, demonstrating their location in $ \mathbb R^4$, 
and the overall factor $ (x_1-x_2 )^{ -2m - 2 n } $. We have chosen 
to keep the notation simple, the position dependences can be added
 back easily if desired. 
Likewise, analogously to (\ref{cprojorth}) we have 
 for the correlator of central projectors 
\bea\label{cprojorthft} 
&& < 0  | :  tr _{m,n} ( \Sigma ( P^{\gamma_2 } ) ) 
(   \Phi^{\dagger }  \otimes  \Phi ) : ~ :
 tr_{m,n}  (   \Sigma ( P^{\gamma_1 } )  
 (  \Phi \otimes  \Phi^{\dagger} ) ) :  | 0 > \cr 
&& = m ! n ! \delta_{ \gamma_1 \gamma_2 } tr_{m,n}  ( P^{\gamma_1}   ) \cr 
&& = m! n! \delta_{ \gamma_1 \gamma_2 } ~ d^{(B)}_{\gamma_1} ~ Dim \gamma_1  
\eea
The special property of the $k=0$ projectors is that they 
are orthogonal to contractions $ C_{ i \bar j }$. 
In the special case of $ k=0 $, and  denoting 
$ \gamma_+ = R , \gamma_- = S $,  
\bea 
C_{ i \bar j } P_{ R \bar S } = 0 
\eea 
This means that the corresponding operators have no short 
distance singularities 
\bea  
: tr _{m,n} (   \Sigma ( P^{\gamma_1 } )  
 (  \Phi \otimes  \Phi^{\dagger} ) ) :
~ = ~   tr_{m,n}  (   \Sigma ( P^{\gamma_1 } )  
 (  \Phi \otimes  \Phi^{\dagger} ) )
\eea 

Any $k=0 $ irrep of Brauer determines a pair of Young diagrams 
$ ( R , S ) $ and a projector $ P_{ R \bar S } $. The nonsingular 
field  theory 
operator $ tr P_{ R \bar S } ( \Phi \otimes \Phi^{\dagger} ) =   
 tr \Sigma ( P_{ R \bar S }  ) ( \Phi \otimes \Phi^{\dagger} )
 $ 
is our proposal for a giant-anti-giant operator where 
$R$ determines a giant and $S$ determines an anti-giant. 
These operators only exist when $ c_1 ( R ) + c_1 ( S ) \le N $, 
as is clear from (\ref{nonchirschwe}). We will describe this cutoff 
as a non-chiral stringy exclusion principle, and we will discuss it further in 
section \ref{sec:finiteN}. In the chiral case the proposal reduces 
to $ tr ( P_R \Phi ) = d_R \chi_R ( \Phi ) $. 
For $k \ne 0 $ operators we expect that the subtraction will 
involve powers up to $ \epsilon^{-2k} $ in the
short distance subtraction.

\subsection{ Examples : $B_{N} ( 3,1 )$ and $ B_{N} (2,2 )$ }
We consider here some simple examples of $B_{N}(m,n)$ 
with $m+n<N$. 
Take for example $m=3$, $n=1$. In this case 
there are 
$7$ independent gauge invariant operators 
\bea 
&& 
( tr \Phi )^3 tr  \Phi ^{\dagger }  
\qquad tr   \Phi ^3   tr  \Phi ^{\dagger } 
\qquad
tr \Phi ^2 tr  \Phi  tr \Phi ^{\dagger}
\cr
&& 
tr   \Phi ^3 \Phi ^{\dagger }   
\qquad 
tr   \Phi ^2 \Phi ^{\dagger }  tr   \Phi  
\qquad 
   tr \Phi ^2 tr  \Phi \Phi ^{\dagger} 
\qquad   tr \Phi  \Phi ^{\dagger }  ( tr \Phi  )^2 
\eea 
In the case of $m=2$, $n=2$, we have $10$ 
gauge invariant operators 
\begin{eqnarray}
&&
tr  \Phi^2 tr (\Phi^{\dagger })^{2}  
\qquad 
(tr  \Phi)^{2} tr (\Phi^{\dagger })^{2}  
\qquad 
tr  \Phi^2 (tr \Phi^{\dagger })^{2}  
\qquad 
(tr  \Phi)^{2}(tr \Phi^{\dagger })^{2}  
\cr
&&
tr  \Phi^2 \Phi^{\dagger } tr \Phi^{\dagger }
\qquad 
tr  \Phi tr \Phi(\Phi^{\dagger })^{2}  
\qquad 
tr \Phi tr\Phi \Phi^{\dagger }tr \Phi^{\dagger }
\qquad 
(tr\Phi \Phi^{\dagger })^{2}
 \cr
&&
tr\Phi^2(\Phi^{\dagger })^2
\qquad 
tr\Phi \Phi^{\dagger }\Phi \Phi^{\dagger }
\end{eqnarray}
In both cases, these listed operators do not form 
orthogonal bases. 
A complete set of orthogonal bases of gauge invariant operators 
is obtained by taking linear combinations of 
these operators, and finding such 
a set is solved by the use of 
symmetric projectors. 
It is explicitly shown in 
appendix \ref{symmetricoperators}. 
By making the following replacement 
\begin{equation}
\Phi \rightarrow A^{\dagger}, \quad 
\Phi^{\dagger} \rightarrow B^{\dagger}
\end{equation} 
we obtain a complete set of orthogonal gauge invariant states 
in the quantum mechanics.

\section{ Physical interpretation of  the spectrum of multi-traces } 
\label{sec:finiteN}

We have provided  a basis of multi-trace 
constructed from $ \Phi, \Phi^{\dagger}  $ which diagonalise the two-point 
function in free  4D $N=4$ SYM (\ref{brauerdiagcorft}), by using the 
symmetric branching operators $ Q^{\gamma }_{A,  i j }$. 
By the operator-state correspondence, these give rise to orthogonal states. 
The symmetric branching operators 
 also give an orthogonal basis of states in the reduced 
quantum mechanics  of two matrices (\ref{pfdiag}).
 The same combinations of multi-traces also give diagonal correlators 
 in the zero-dimensional matrix model (\ref{zeropart}).

The label $\gamma $ in (\ref{nonchirschwe}) 
 identifies  simultaneously the irreps of 
$U(N)$ and $B_N(m,n)$ which appear in the decomposition 
of $\vmn$.  
 In the Young diagram 
description of $U(N)$ negative row lengths are allowed. 
The set of positive rows defines $\gamma_+$ and the set of negative 
rows define $ \gamma_- $. For example with $m = 6, n=4 , N = 6 $ a 
possible $ \gamma $ is  $ \gamma = [ 3,2,1,  -1, -1 , -2 ] $. 
This determines $ \gamma_+ = [ 3,2,1] $ , $ \gamma_- = [ 2,1,1] $ 
and $ k=0 $. An example such as  
$ \gamma = [ 2,1,1,  1, -1 , -2 ] $
 determines   $ k =1 ,\gamma_+ = [2,1,1,1] , \gamma_- = [ 2,1 ] $. 
Given a pair of Young diagrams $R, S $ with 
$m,n$ boxes respectively, and $ c_1 ( R ) + c_1 ( S ) \le N $, 
there is always a $\gamma$ with $k=0$ 
\bea\label{compos}  
&& \gamma_i = r_i   ~~~ \hbox{for } i = 1 \cdots c_1 \cr 
&& \gamma_i = 0 ~~~ \hbox{for }i = c_1 +1 ... N-\bar c_1 \cr 
&& \gamma_i  = -s_{N-i+1} ~~~  \hbox{for } i = N - \bar c_1+1 \cdots N 
\eea 
This $\gamma$ corresponds to  $ k=0 , \gamma_+ = R , \gamma_- = S $. 
For such a $\gamma$, the $Q^{\gamma}_{A ,  i j} $ reduces to 
a single central projector, which we have called $P_{R \bar S } $. 
The multi-trace operator  associated to this projector is 
our proposal for the ground state of the brane-anti-brane system
made as a composite of the brane described by $R$ and the 
anti-brane described by $S$.
The $k=0$ projectors are annihilated by the contraction 
operators in $B_N(m,n ) $ as a result they have no short distance 
singularities.  As explained in section 2, the Wick-contractions 
 between $ \Phi $ and $ \Phi^*$ result in  a 
contraction in tensor space. When we construct a composite  
 operator by  composing a projector $P_{ R \bar S } $ 
with  $ \Phi \otimes \Phi^* $ in $ \vmn $ 
i.e $tr_{m,n} ( P_{ R \bar S }  \Phi \otimes \Phi^* ) $,  and we consider 
the short distance subtractions we encounter precisely the 
products $C_{i \bar j } P_{ R \bar S } = 0$. Hence the short-distance 
singularities in $tr_{m,n} ( P_{ R \bar S }  \Phi \otimes \Phi^* ) $
vanish without the need for subtractions. Equivalently 
\bea 
: ~ tr_{m,n} ( P_{ R \bar S }  \Phi \otimes \Phi^*)~ : ~ = ~  
tr_{m,n} ( P_{ R \bar S }  \Phi \otimes \Phi^* )
\eea 
The construction of a composite $\gamma$ from the 
pair $ R , S $ as in (\ref{compos}) is nothing but the addition 
of the  $ U(N)$ weights associated with $R$ and $\bar S $, 
which also appears in 2dYM. The $U(N)$ weights are related to 
 generalised spacetime charges connected  to the 
integrability of the theory \cite{bclj} 
and from this point of view it may be possible to develop a  
spacetime interpretation of  the addition of weights  for the composite 
system.

 The critical reader might object that we should 
 not expect an exact eigenstate of the string theory Hamiltonian corresponding 
to a brane-anti-brane configuration  which should be unstable due to tachyon 
condensation.  However, at zero Yang-Mills coupling, or the tensionless string 
limit, the tachyon formally becomes massless. So we should expect 
a map between brane-anti-brane configurations and exact
 eigenstates to be possible. Even in the opposite limit of
 semiclassical gravity, with a probe description of giants and anti-giants
following \cite{mst} we might expect that when the brane and anti-brane 
are far from each other there should be at least an approximate eigenstate 
corresponding to their composite. In fact the description of 
static non-supersymmetric supergravity solutions in terms of brane-anti-brane 
parameters ( for a recent discussion and references to the relevant literature 
see \cite{staticddbar} )  suggests that it might also be possible to 
extrapolate an appropriate brane-anti-brane description of the eigenstates 
of the free Yang-Mills Hamiltonian to the supergravity regime, where 
 back-reaction on spacetime  would produce a non-supersymmetric 
generalization of the solutions of LLM \cite{llm}.   

The multi-traces constructed from $P_{R \bar S } $ thus 
give us the lowest energy state we can associate to the 
composite of brane $R$ and anti-brane $ S $.
 For the more general $Q$ operators, 
associated with  $ \gamma ( k \ge  1 )$, a natural proposal is that 
for $\gamma=  ( k , \gamma_+ \vdash m-k , \gamma_- \vdash n-k )$ we have states 
of energy $m+n $, which are excitations of the brane-anti-brane 
system $ ( \gamma_+ , \gamma_- )$. The excitations carry $2k$ units 
of energy and have a multiplicity 
\bea 
\sum_{ R \vdash m , S \vdash n  }  ( M^{\gamma}_{RS } )^2 
\eea
This multiplicity and the 
 form of the corresponding  operators $Q^{\gamma}_{ (R  S) ,   i  j } $  
suggest that they should also be interpreted as 
states arising from a descent procedure 
 from the brane-anti-brane pair $ R , S $ of energies $m,n$. 
The descent procedure in question involves partons 
( constituents of the brane each carrying a unit of angular momentum
each ) and anti-partons ( constituents of the anti-brane each carrying a 
 unit of angular momentum ). The integer $k$ can be interpreted 
as the number of partons and anti-partons, from the brane-anti-brane 
pair $R,S $  combining to form a stringy excitation  
with $2k$ units of energy. For fixed initial brane-configuration $R, S $ 
and final brane configuration $( \gamma_+ , \gamma_- ) $ 
the multiplicity $ ( M^{\gamma}_{RS } )^2 $ and the matrix structure of 
the associated operators  $Q^{\gamma}_{ (RS) ,   i j } $ (\ref{branchmatmult})
is suggestive of a Chan-Paton interpretation of the $i,j$ 
indices labelling the  stringy excitations. It would be very interesting to 
construct a dynamical model of the stringy excitations and the descent
procedure which precisely accounts for the multiplicity $ M^{\gamma}_{RS } $
 known in terms of Littlewood-Richardson coefficients (\ref{redcoeff}). 
A useful analogy might be the role of the elementary field $\Phi $
as a parton for long strings in the BMN limit \cite{pp},
 which is developed in terms 
of a concrete string bit model in \cite{verlinde}.

The description  of brane-anti-branes in terms of irreps $ \gamma $
of  the Brauer algebras
 $B_N(m,n )$ or $ U(N)$ reveals an interesting finite $N$ cutoff
on  the brane-anti-brane configurations. Let us go back to the 
example of $ m =6, n= 4 , N=6 $. 
Since $\gamma$ has exactly $6 $ rows, no choice of 
$\gamma $ will give a pair such as 
$ \gamma_+ = [ 2,2,1,1] , \gamma_- = [2,1,1]$. In general we 
have the bound 
\bea\label{ncsep} 
 c_1 ( \gamma_+ ) + c_1 ( \gamma_- ) \le N 
\eea 
We will call this the {\bf non-chiral stringy exclusion principle}, 
following the terminology of stringy exclusion principle 
for the cutoff in the holomorphic case \cite{malstrom}.  
The bound on the individual branes or anti-branes 
 $ c_1 ( \gamma_+ ) \le  N $ and  $c_1 ( \gamma_- ) \le  N $
can be understood in the semiclassical probe picture \cite{mst} 
or the supergravity picture \cite{llm}  and a more speculative
 explanation of related cutoffs on single traces 
in terms  of non-commutative spacetime
 was  explored \cite{jr}.
 An analogous spacetime understanding of the  the new bound 
on composites 
is a fascinating challenge for the spacetime picture. 
If we are given a pair $ \gamma_+ = R  , \gamma_- = S $ violating the bound we 
can still form a $U(N)$ irrep $ \gamma $ by adding the
 corresponding highest weights as follows 
\bea\label{gammaRSE}  
&& \gamma_{i} = r_i ~~~ \hbox{ for} ~ i =1 \cdots c_1 -E \cr 
&& \gamma_i = r_i - s_{N-i+1}~~~  \hbox{ for }~ i = c_1 - E +1 \cdots c_1 \cr 
&& \gamma_i = - s_{N-i+1}  ~~~ \hbox{for } ~ i = c_1 +1 \cdots N 
\eea 
It is  easy to check that $ \gamma_{i} \ge \gamma_{i+1} $  as 
required. This $\gamma $ has  
$ k = \sum_{i=c_1 - E +1 }^{c_1} min ( r_i , s_{N-i+1} ) $, 
so that $2k$ is the number of boxes removed after superposing the 
Young diagram of $ R$ with that of $ \bar S $ to get the $ \gamma$. 
Following the interpretation above, this says that 
when the bound is violated, the composite brane-anti-brane system 
is actually   a stringy excited state of a  brane-anti-brane pair 
$ ( \gamma_+ , \gamma_- )  $ determined by (\ref{gammaRSE}) with energy 
$ m +n - 2 k =  m+n -2 \sum_{i=c_1 - E +1 }^{c_1} min ( r_i , s_{N-i+1} ) $, 
which has been excited by a stringy excitation with $2k$ units of energy.

We will now describe some technical consequences of the 
non-chiral stringy exclusion principle. 
Since pairs $\gamma = ( k= 0,  R \vdash m  , S \vdash n )  $
 do not appear in (\ref{nonchirschwe}) 
 when $c_1 ( R ) + c_1 ( S ) > N $, we may ask what happens to 
 the general formulae for $P_{ R \bar S } $ we wrote in 
this case. It turns out that in some cases, the projector 
becomes ill-defined with $0$ appearing in the denominator. 
In other cases, it can be seen that the projector 
acting on $ \vmn$ gives zero. This means 
that $ tr_{m,n} ( P_{R \bar S } \Phi \otimes \Phi^* ) $ vanishes. 
Since the first term in $  P_{R \bar S } $ is 
$ p_R p_S $, this means that the vanishing 
leads to a matrix identity for $ \chi_R ( \Phi  ) \chi_S ( \Phi^{\dagger}  ) $ 
in terms of multi-traces where $ \Phi , \Phi^{\dagger} $
 appear in the same trace.

Before getting to a non-chiral example we review 
an analog  in the chiral case. Matrix identities follow from the 
fact that the $n$-fold  antisymmetiser acting on the $N$-dimensional 
space $V$ vanishes when $ n > N $. 
\begin{eqnarray}
p_{[1^{n}]}V^{\otimes n}=0 \quad (n>N)
\label{chiralPV=0}
\end{eqnarray}
Let us consider an $N=2$ matrix $\phi$. 
In this case, we have the following identity
\begin{eqnarray}
tr(\phi^{3})
=
\frac{3}{2}
(tr\phi )tr(\phi^{2})-\frac{1}{2}(tr\phi)^{3}
\label{matrixidentityforN=2}
\end{eqnarray}
which is a direct consequence of (\ref{chiralPV=0}) because 
(\ref{matrixidentityforN=2}) can be rewritten as 
\begin{eqnarray}
tr_{3}(p_{[1^{3}]}\phi)
=
\frac{1}{6}\left(
(tr\phi)^{3}-3(tr\phi) tr(\phi^{2})+2tr(\phi^{3})
\right)
=0
\end{eqnarray}

Now we consider a non-chiral example of matrix identities 
following from the vanishing of projectors. 
For example, we can show 
\begin{eqnarray}
P_{[1^{2}]\bar{[1]}} V^{\otimes 2}\otimes \bar{V}  = 0
\quad (N=2) 
\label{PVV=0N=2}
\end{eqnarray}
as follows.  
For a state $w=v_{1}\otimes  v_{2}\otimes \bar{v}_{1} \in 
V^{\otimes 2}\otimes \bar{V}$, 
we have 
\begin{eqnarray}
p_{[1^{2}]}w=
v_{[1}\otimes v_{2]} \otimes \bar{v}
\end{eqnarray}
and 
\begin{eqnarray}
p_{[1^{2}]} C w
&=&
\sum_{k=1}^{2}
v_{[k}\otimes v_{2]} \otimes \bar{v}_{k}
\cr
&=&
v_{[1}\otimes v_{2]} \otimes \bar{v}_{1}
\end{eqnarray}
Hence  using  (\ref{b21k0}), with $N=2$ we obtain 
\begin{eqnarray}
P_{[1^{2}]\bar{[1]}}w=\left(1-C\right)p_{[1^{2}]}w=0
\end{eqnarray}
From this equation, 
for two $N=2$ matrices $ A , B $, 
we have 
\begin{eqnarray}
tr_{2,1}  (  P_{[1^{2}]  \bar{[1]}} A\otimes A\otimes B^T ) = 
tr_{2,1}( \Sigma ( P_{[1^{2}]  \bar{[1]}} ) A\otimes A\otimes B   ) =0
\end{eqnarray}
This gives the following matrix identity, 
\begin{eqnarray}
tr_{2}(p_{[1^{2}]}A)tr(B)=tr(A)tr(AB)-tr(A^{2}B)
\end{eqnarray}
where 
\begin{eqnarray}
tr_{2}(p_{[1^{2}]}A)=\frac{1}{2}(trA)^{2}-\frac{1}{2}tr(A^{2}). 
\end{eqnarray}

For two $N=3$ matrices $A$ and $B$, 
we have the following identity  
\begin{eqnarray}
tr_{2}(p_{[1^{2}]}A)tr_{2}(p_{[1^{2}]}B)
&=&trAtrBtrAB-trA^{2}BtrB-trAtrAB^{2}+trA^{2}B^{2}
\cr
&&-\frac{1}{2}trABtrAB+\frac{1}{2}trABAB
\end{eqnarray}
This equation comes from
\begin{eqnarray}
tr_{2,2}  (  P_{[1^{2}] {[\bar 1^{2}]}}     A^{ \otimes 2 } 
\otimes  B^{T~ \otimes 2  } ) 
 =     tr_{2,2}(  \Sigma ( P_{[1^{2}] {[\bar 1^{2}]}} ) A^{ \otimes 2}
\otimes B^{ \otimes  2})=0
\end{eqnarray}
which is a consequence of 
\begin{eqnarray}
P_{[1^{2}][\bar 1^{2} ]} V^{\otimes 2}\otimes \bar{V}^{\otimes 2}  = 0
\quad (N=3)
\end{eqnarray}

The skeptical  reader might wonder  if we can 
bypass the nonchiral exclusion principle (\ref{ncsep}) simply 
by proposing $ : \chi_R ( \Phi ) \chi_S ( \Phi^{\dagger}  ) : $ 
as a dual. The first objection to this is that such a proposal 
does not belong to a diagonal basis. The $P_{ R \bar S } $ 
are a special case of the complete orthogonal set of 
symmetric branching operators 
of section 7. But a more startling failure  of the naive proposal 
is the fact explained above, that  at finite $N$,  the products of characters 
become equal to sums of multi-traces  where products $ \Phi \Phi^{\dagger} $ 
appear within the same trace.

\section{ Summary and Discussion }

For any Young diagram $R$, there is an operator
$ \chi_R ( \Phi )$ and a  spherical D-brane giant graviton. 
For any Young diagram $S$, there is a spherical $\bar D3$-brane 
giant graviton and corresponding operator $ \chi_S ( \Phi^{\dagger } )$.
$ \chi_R ( \Phi ) $ is a holomorphic continuation 
of characters  $\chi_R ( U   ) $ of the unitary group. 
We can view $ \chi_S ( \Phi^{\dagger}  ) $ is a continuation 
of  $ \chi_S ( U^{\dagger}  ) $. Associated with a pair of Young diagrams, 
are ``coupled representations'' of $U(N)$ which play an important 
role in 2dYM. The coupled character 
 is obtained from the  trace in $\vmn$   of a  projector (\ref{projcharnc}).  
 These projectors are constructed from Brauer 
algebras $B_N ( m,n ) $. 
 These same projectors can be used to construct composite 
operators involving the complex matrix $ \Phi $ and its conjugate $ \Phi^* $. 
We have given a number of useful general formulae for these operators 
in section 4, and discussed several examples in section 5 and the appendices. 
These operators are proposed as candidate gauge theory duals 
 for brane-anti-brane systems. They exist when 
$ c_1 ( R ) + c_1 ( S ) \le N $. 
They have an interesting property that 
they are unchanged by short distance subtractions, generalising the 
simple example discussed in section 2. 
We also described the complete set of operators 
that can be constructed from $ \Phi , \Phi^{\dagger}  $. 
They are related to symmetric elements of $ B_{N} ( m , n )$, 
i.e those invariant under conjugation by the subalgebra 
$ S_m \times S_n $. We gave a group theoretic counting of these 
operators and described corresponding branching operators 
which lead to an orthogonal basis for the  two point functions. 

Our calculations have been done  in the zero coupling limit $ g_{YM}^2 = 0 $. 
In the case of BPS objects, these results can be extrapolated to
 strong coupling using non-renormalisation theorems, where 
they apply to the weakly coupled gravity regime.  In this 
non-supersymmetric set-up,  important physics should 
be contained in the mixing with  fields other than  
$ \Phi , \Phi^{\dagger } $ which occurs when the coupling is 
turned on. The basis of operators 
described here should be a natural starting point for perturbation away 
from zero coupling. It will be interesting to apply the technology 
for constructing operators corresponding to strings between branes, 
developed in \cite{Balasubramanian:2004nb,
de Mello Koch:2007uu,de Mello Koch:2007uv} to study the 
strings between brane and anti-brane using our proposed brane-anti-brane 
operators $tr_{m,n} (P_{R \bar S } \Phi \otimes \Phi^* ) $. 
This should shed light on 
tachyon condensation 
(see \cite{Sen:1999mg,Sen:2004nf} for reviews) 
in AdS from the dual gauge theory point of view, in regimes 
not  accessible to perturbative string theory. 

We have related the counting of multi-trace operators to 
Brauer algebras and using this interpretation of the counting 
we have proposed a brane-anti-brane interpretation of the operators
(section 8). We uncovered a non-chiral stringy exclusion principle 
(\ref{ncsep}) :  a cut-off on brane-anti-brane states which is stronger 
than the cutoff on the individual branes and anti-branes. 
 These general  lessons on  the counting and  finite $N$ effects  
should be relevant at strong coupling.

The AdS/CFT set-up thus has allowed the unique opportunity 
to find exact quantum operators corresponding to brane-anti-branes. 
It is natural to ask if there are lessons here for 
  brane-anti-brane physics more generally.  
 Brane-anti-brane degrees of freedom are used in black-hole counting. 
   It has been generally a confusing issue, whether we can expect 
   these unstable configurations to correspond to exact eigenstates of a
 String Theory Hamiltonian. The lesson here is that we can certainly 
 expect exact eigenstates because we are taking a limit 
 of zero tension, where the tachyon becomes massless. In this 
 limit we have been able to construct exact
 operators for brane-anti-brane systems using Brauer algebras, 
and classified the stringy excitations of the brane-anti-brane systems 
using these algebras. It will be interesting to see how far one can extend 
 this discussion to more general backgrounds and to the counting 
of the degrees of freedom of stringy non-supersymmetric systems 
such as those considered in \cite{hms}.

 One of our main objects of interest in this paper 
 has  been about projectors in the Brauer algebra.
 Another algebraic structure which captures
 many properties of tachyon  condensation
is K-theory \cite{moomin,witten2,olszab}. The algebraic version of K-theory 
 involves equivalence classes of projectors.
It is interesting to ask if the Brauer algebras, 
 perhaps in an inductive limit of $ m , n \rightarrow \infty $, 
 and their connections to K-theory,    might provide an
 algebraic structure which is relevant 
 to brane-anti-brane systems in a general background.

\newpage 

\vskip1in 

{ \Large 
{ \centerline { \bf Acknowledgements }  } } 

\vskip.2in 
We thank  Tom Brown, Paul Heslop, Hally Ingram,  Horatiu Nastase, 
Costis Papageorgakis, Gabriele Travaglini for useful discussions. SR is 
supported by a  STFC Advanced Fellowship and in part by the
 EC Marie Curie Research Training Network MRTN-CT-2004-512194. 
 YK is supported by STFC grant PP/D507323/1 
 ``String theory, gauge theory and gravity''. 
He thanks theoretical physics laboratory at RIKEN for hospitality 
while part of this work was done 
and Yuuichirou Shibusa for useful discussions. 
He also thanks the Yukawa Institute for Theoretical Physics 
at Kyoto University. 
Discussions during the YITP workshop YITP-W-07-05 on 
``String Theory and Quantum Field Theory'' were useful 
to complete this work.

\vskip.5in 

\begin{appendix}

\section{Appendix   }\label{appendixproj1}

\subsection{Calculation of duals of Brauer elements }
\label{sec:calculationof1ast}
In this section, we consider the duals of Brauer elements. 
In particular we use (\ref{1ast}) to find explicit formulae 
for the dual of the identity 
$1^{\ast}$  for some examples.  We use this to 
 obtain projectors for $k=0$ representations 
using the formula (\ref{rewrittenprojector}). 
In the calculation of $1^{\ast}$, we need values of the 
character of the symmetric group. 
They are listed in \cite{fulhar}, for example. 

\subsubsection{Duals for $B_N(1,1)$}
$B_N(1,1)$ can be mapped to $\mathbb{C }[S_{2}]$. 
The $\Omega_{2}$ factor in $S_{2}$ is given by 
$\Omega_{2}=1+s_{1}/N$, and the inverse of it is 
\begin{equation}
\Omega_{2}^{-1}=\frac{N^{2}}{N^{2}-1}\left(1-\frac{s_{1}}{N}\right)
\end{equation}
Because the inverse map $\Sigma^{-1}$ of $s_{1}$ is given by 
$C_{1\bar{1}}$, 
$1^{\ast}$ can be calculated as 
\begin{eqnarray}
1^{\ast}=\frac{1}{N^{2}-1}
\left(
1-\frac{C_{1\bar{1}}}{N}
\right) 
\end{eqnarray}
Then using 
\begin{eqnarray}
t_{[1]\bar{[1]}}=N^{2}-1
\end{eqnarray}
we get a $k=0$ projector 
\begin{eqnarray}
P_{[1][1]}
=
t_{\lambda}
1^{\ast}=1-\frac{C_{1\bar{1}}}{N}
\end{eqnarray}
which can be easily checked to satisfy 
$(P_{[1][1]})^{2}=P_{[1][1]}$ using 
$(C_{1\bar{1}})^{2}=NC_{1\bar{1}}$.  

\subsubsection{Duals for $B_N(2,1)$}
In this case, $1^{\ast}$ is given by 
\begin{eqnarray}
\frac{1}{\left(3!\right)^{2}}
\sum_{ T }\frac{d_{T}^{2}}{dim T }
\sum_{\sigma\in S_{3}}\chi_{T }(\sigma^{-1})
\Sigma^{-1}
\left( \sigma \right)
\end{eqnarray}
where $T$ is an irrep of $S_{3}$. 
The relationship between $B_N(2,1)$ and $\mathbb{C }[S_{3}]$ is 
\begin{eqnarray}
&&\Sigma^{-1}(T_{[2,1]})=s_{1}+C \cr
&&\Sigma^{-1}(T_{[3]})=Cs_{1} 
\end{eqnarray}
where $T_{r}$ is the sum of elements belonging to a conjugacy class 
$r$ which is labelled by the cycle structure.  
Using the values of the character\footnote{Note that 
$\chi_{R}(g)=\chi_{R}(g^{\prime})$ when $g$ and $g^{\prime}$ 
are conjugate each other. } 
\begin{eqnarray}
&&\chi_{[3]}(1)=d_{[3]}=1 \quad 
\chi_{[3]}(s_{1})=1 \quad \chi_{[3]}(s_{1}s_{2})=1 \cr
&&\chi_{[1^{3}]}(1)=d_{[1^{3}]}=1\quad 
\chi_{[1^{3}]}(s_{1})=-1 \quad  \chi_{[1^{3}]}(s_{1}s_{2})=1 \cr
&&\chi_{[2,1]}(1)=d_{[2,1]}=2 \quad 
\chi_{[1^{3}]}(s_{1})=0 \quad  \chi_{[1^{3}]}(s_{1}s_{2})=-1 
\label{characterS3}
\end{eqnarray}
$1^{\ast}$ can be calculate 
as 
\begin{eqnarray}
1^{\ast}=\frac{1}{N(N-s)(N+2s)}
\left(
1-\frac{1}{N\Omega_{2}}C
\right) 
\end{eqnarray}
Using 
\begin{eqnarray}
&&t_{[2]\bar{[1]}}=\frac{1}{2}N(N-1)(N+2)\cr
&&t_{[1,1]\bar{[1]}}=\frac{1}{2}N(N+1)(N-2)
\end{eqnarray}
we obtain two $k=0$ projectors corresponding to 
$R=[2]$ or $[1^{2}]$
\begin{eqnarray}
P_{R\bar{[1]}}
&=&
t_{R\bar{[1]}}
1^{\ast}
\frac{m!}{d_{R}}p_{R} \cr
&=&
t_{R\bar{[1]}}
\frac{1}{N}\frac{1}{(N-s)(N+2s)}
\left(
1-\frac{1}{N\Omega_{2}}C
\right)2p_{R} \cr
&=&
\left(
1-\frac{1}{N\Omega_{2}}C
\right)p_{R} 
\end{eqnarray}
where we have used $sp_{[2]}=p_{[2]}$ and $sp_{[1,1]}=-p_{[1,1]}$.

\subsubsection{Duals for $B_N(2,2)$}
In this case, $1^{\ast}$ is given by 
\begin{eqnarray}
\frac{1}{\left(4!\right)^{2}}
\sum_{ T }\frac{d_{T}^{2}}{dim T }
\sum_{\sigma\in S_{4}}\chi_{T }(\sigma^{-1})
\Sigma^{-1}
\left( \sigma \right)
\end{eqnarray}
where $T$ is an irrep of $S_{4}$. 
After some calculations using the values of the character 
and the mapping rule 
\begin{eqnarray}
&& \Sigma^{-1}(T_{[2,1^{2}]})=C+s+\bar{s} \cr
&& \Sigma^{-1}(T_{[3,1]})=C(s+\bar{s}) \cr
&& \Sigma^{-1}(T_{[4]})=Cs\bar{s}+C^{(2)}s \cr
&& \Sigma^{-1}(T_{[2,2]})=C^{(2)}+s\bar{s} 
\end{eqnarray}
we get the following expression 
\begin{eqnarray}
1^{\ast}
&=&
\frac{1}{N^{2}(N^{2}-1)(N^{2}-4)(N^{2}-9)}\times  \cr
&&
\Big(N^{4}-8N^{2}+6-N(N^{2}-4)\Sigma^{-1}(T_{[2,1^{2}]})
+(2N^{2}-3)\Sigma^{-1}(T_{[3,1]}) \cr
&&
-5N\Sigma^{-1}(T_{[4]})+(N^{2}+6)\Sigma^{-1}(T_{[2,2]})
\Big) 
\label{1^{ast}ofS_{4}}
\\
&=&
\frac{1}{N^{2}(N^{2}-1)(N^{2}-4)(N^{2}-9)}\times  \cr
&&
\Big(N^{4}-8N^{2}+6-N(N^{2}-4)(C+s+\bar{s})
+(2N^{2}-3)C(s+\bar{s}) \cr
&&
-5N(Cs\bar{s}+C^{(2)}s)+(N^{2}+6)(C^{(2)}+s\bar{s})
\Big)
\end{eqnarray}
where $C=C_{1\bar{1}}+C_{1\bar{2}}+C_{2\bar{1}}+C_{2\bar{2}}$ 
and 
$C_{(2)}=C_{1\bar{1}}C_{2\bar{2}}+C_{1\bar{2}}C_{2\bar{1}}$.

We consider $(R,S)=([2],[2])$. 
The use of $sp_{[2]}=p_{[2]}$ and 
$\bar{s}\bar{p}_{[2]}=\bar{p}_{[2]}$ simplifies the expression 
of $1^{\ast}p_{[2]}\bar{p}_{[2]}$ as 
\begin{eqnarray}
1^{\ast}p_{[2]}\bar{p}_{[2]}
=\frac{1}{N^{2}(N-1)(N+3)}\left(
1-\frac{1}{N+2}C+\frac{1}{(N+1)(N+2)}C_{(2)}
\right)
\end{eqnarray}
Using 
\begin{eqnarray}
t_{[2]\bar{[2]}}
=\frac{1}{4}N^{2}(N-1)(N+3)
\end{eqnarray}
we obtain a projector corresponding to $(R,S)=([2],[2])$
\begin{eqnarray}
P_{[2]\bar{[1]}}=\left(
1-\frac{1}{N+2}C+\frac{1}{(N+1)(N+2)}C_{(2)}
\right)p_{[2]}\bar{p}_{[2]}
\end{eqnarray}
Other cases 
$(R,S)=([1^{2}],[1^{2}]),([2],[1^{2}]),([1^{2}],[2]])$
can be done similarly to obtain (\ref{t=0form2n2}). 


\subsubsection{Duals for $B_N(3,1)$}
Because $B_N(3,1)$ can also be mapped to $\mathbb{C }[S_{4}]$, 
we can use the equation (\ref{1^{ast}ofS_{4}}). 
Though both of these two cases $B_N(2,2)$ and $B_N(3,1)$
can be mapped to the same group algebra $\mathbb{C }[S_{4}]$, 
the mapping rule is different, and the mapping rule in this 
case is given by 

\begin{eqnarray}
&& \Sigma^{-1}(T_{[2,1^{2}]})=C+T_{[2,1]} \cr
&& \Sigma^{-1}(T_{[3,1]})=CT_{[2,1]}-D+T_{[3]} \cr
&& \Sigma^{-1}(T_{[4]})=CT_{[3]} \cr
&& \Sigma^{-1}(T_{[2,2]})=D
\label{mapB(31)S4}
\end{eqnarray}
where 
$C=C_{1\bar{1}}+C_{2\bar{1}}+C_{3\bar{1}}$, 
$D=s_{2}C_{1\bar{1}}+
s_{1}s_{2}s_{1}C_{2\bar{1}}+s_{1}C_{3\bar{1}}$. 
Note that $T$ in the LHS of (\ref{mapB(31)S4}) 
is an element of $S_{4}$ while 
$T$ in the RHS is an element of $S_{3}$. 
Then we obtain
\begin{eqnarray}
1^{\ast}
&=&
\frac{1}{N^{2}(N^{2}-1)(N^{2}-4)(N^{2}-9)}\times  \cr
&&
\Big(N^{4}-8N^{2}+6-N(N^{2}-4)(C+T_{[2,1]})
+(2N^{2}-3)(CT_{[2,1]}-D+T_{[3]}) \cr
&&
-5N(CT_{[3]})+(N^{2}+6)(s_{2}C_{1\bar{1}}+
s_{1}s_{2}s_{1}C_{2\bar{1}}+s_{1}C_{3\bar{1}})
\Big)
\end{eqnarray}
Note that $T$ is a central element, and 
using $T_{r}p_{R}=(\chi_{R}(T_{r})/d_{R})p_{R}$, 
we get 
\begin{eqnarray}
&&1^{\ast}p_{[3]}
=\frac{1}{N(N^{2}-1)(N+3) }\left(
1-\frac{1}{N+2}C \right)
\cr
&&1^{\ast}p_{[1^{3}]}
=\frac{1}{N(N^{2}-1)(N-3) }\left(
1-\frac{1}{N-2}C \right)
\cr
&&1^{\ast}p_{[2.1]}
=\frac{1}{N^{2}(N^{2}-4)}\left(
1-\frac{N}{N^{2}-1}\Omega_{2}^{<i>}C_{i\bar{1}} \right)p_{[2.1]}
\label{1astforB31}
\end{eqnarray}
where we have used 
$s_{i}p_{[3]}=p_{[3]}$, $s_{i}p_{[1^{3}]}=-p_{[1^{3}]}$ and 
the values of the character (\ref{characterS3}). 
Using 
\begin{eqnarray}
&&t_{[3]\bar{[1]}}=\frac{1}{6}N(N^{2}-1)(N+3) \cr
&&t_{[1^{3}]\bar{[1]}}=\frac{1}{6}N(N^{2}-1)(N-3) \cr
&&t_{[2,1]\bar{[1]}}=\frac{1}{3}N^{2}(N^{2}-1)(N+3) 
\end{eqnarray}
the three equations (\ref{1astforB31}) 
are unified to be 
\begin{eqnarray}
1^{\ast}p_{R}
=\frac{d_{R}}{3!t_{R\bar{S}}}\left(
1-\frac{1}{N\Omega_{3}}\Omega_{2}^{<i>}C_{i\bar{1}} \right)p_{R}
\end{eqnarray}
and we reproduce the expression of projectors (\ref{t=0m3n1}).


\subsection{ Algebra relations in  $B_N(3,1)$ and $B_N(2,2) $ } 
In this section, we list some useful formulae for 
$B_N(3,1) $ and $B_N(2,2) $ which are used in the construction 
of projectors in sections \ref{b31proj} and \ref{b22proj}.  
\subsubsection{$B_N(3,1) $}
\label{sec:appB(31)}
\begin{eqnarray}
&& C^{2}=NC+T_{[2,1]}C- D \cr
&& D^{2}=NC+T_{[2,1]}C-D \cr
&& CD=ND+T_{[3]}C
\end{eqnarray}
where 
$T_{[2,1]}=s_{1}+s_{2}+s_{1}s_{2}s_{1}$ and 
$T_{[3]}=s_{1}s_{2}+s_{2}s_{1}$ are central elements in 
$S_{3}$. 
\begin{eqnarray}
(T_{[3]})^{2}=T_{[3]}+2, \quad (T_{[2,1]})^{2}=3(T_{[3]}+1), \quad
T_{[2,1]}T_{[3]}=2T_{[2,1]}
\end{eqnarray}
\begin{eqnarray}
&&T_{[2,1]}D=C+T_{[3]}C \cr
&&T_{[3]}D=T_{[2,1]}C-D 
\end{eqnarray}

\subsubsection{$B_N(2,2) $}
\label{sec:projector22}
\begin{eqnarray}
&&C_{2}=(N+s+\bar{s})C+2C_{(2)} \cr
&&CC_{(2)}=C_{(2)}C=C_{(2)}(2N+s+\bar{s}) \cr
&&(C_{(2)})^{2}=C_{(2)}N(N+s) \cr
&&C_{(2)}s=C_{(2)}\bar{s}
\end{eqnarray}

\subsection{Proof of $C_{i\bar{j}}P_{R\bar{S}}=0$ for 
composites of symmetric and antisymmetric representations}
\label{sec:CP=0forsymandantisym}
In this section, we give a proof of $C_{i\bar{j}}P_{R\bar{S}}=0$ 
for $R=[m]$ or $[1^{m}]$ 
and $S=[n]$ or $[1^{n}]$. 

We first consider $(R,S)=([m],[n])$.
We recall the expression of the projector in this case 
\begin{eqnarray}
P_{[m]\bar{[n]}}
&=&
\left(1-\frac{1}{N+m+n-2}C_{(1)} \right.\cr
&& \left. \qquad
+\frac{1}{(N+m+n-3)(N+m+n-2)}C_{(2)}
+\cdots
\right)
p_{[m]}\bar{p}_{[n]} \cr
&=&
\left(
1+\sum_{k=1}^{n}(-1)^{k}\prod_{l=1}^{k}
\frac{1}{(N+m+n-l-1)}C_{(k)}
\right)
p_{[m]}\bar{p}_{[n]}
\end{eqnarray}
where 
\begin{eqnarray}
&&
C_{(1)}=\sum_{ij}C_{i\bar{j}} \quad
C_{(2)}=\frac{1}{2}\sum_{i\neq k}\sum_{j\neq l}C_{i\bar{k}}C_{j\bar{l}} 
\quad
\cdots 
\cr
&&
C_{(k)}=\frac{1}{k!}
\sum_{i_{a}\neq i_{b}}\sum_{j_{a}\neq j_{b}}
C_{i_{1}\bar{j_{1}}}C_{i_{2}\bar{j_{2}}}\cdots C_{i_{k}\bar{j_{k}}}
\end{eqnarray}
In order to show $C_{1\bar{1}}P_{[m][n]}=0$, 
we need to evaluate $C_{1\bar{1}}C_{(k)}$ acting on $p_{[m]}\bar{p}_{[n]}$. 
The $k=1$ case is calculated as 
\begin{eqnarray}
C_{1\bar{1}}C_{(1)}p_{[m]}\bar{p}_{[n]}
&=&
\left(
NC_{1\bar{1}}+C_{1\bar{1}}\sum_{k\neq 1}(1k)
+C_{1\bar{1}}\sum_{l\neq 1}(\bar{1}\bar{l})
+ C_{1\bar{1}}\sum_{l\neq 1,k\neq1}C_{k\bar{l}}
\right)
p_{[m]}\bar{p}_{[n]} \cr
&=&
\left(C_{1\bar{1}}(N+m+n-2)
+C_{1\bar{1}}C_{(1)}^{<1,\bar{1}>}\right)p_{[m]}\bar{p}_{[n]}
\end{eqnarray}
where we have defined 
\begin{equation}
C_{(1)}^{<1,\bar{1}>}=\sum_{k\neq1,l\neq 1}C_{k\bar{l}}
\end{equation}
which is the sum of single contractions avoiding $\{1,\bar{1}\}$. 
To show the second line, we 
have used 
\begin{equation}
\sigma p_{[m]}=p_{[m]}
\label{sigmaonsymmetricrep}
\end{equation}
for any transposition $\sigma\in S_{m}$.
We next define 
\begin{equation}
C_{(k)}^{<1,\bar{1}>}=\frac{1}{k!}
\sum_{i_{a}\neq i_{b}\neq 1}\sum_{j_{a}\neq j_{b}\neq 1}
C_{i_{1}\bar{j_{1}}}C_{i_{2}\bar{j_{2}}}\cdots C_{i_{k}\bar{j_{k}}}
\end{equation}
for any $k$, which is the sum of $k$ disjoint contractions avoiding 
$\{1,\bar{1}\}$. 
Then we have 
\begin{eqnarray}
&&C_{1\bar{1}}C_{(2)}p_{[m]}\bar{p}_{[n]}
=
\left(
(N+m+n-3)
C_{1\bar{1}}C_{(1)}^{<1,\bar{1}>}
+C_{1\bar{1}}C_{(2)}^{<1,\bar{1}>}\right)
p_{[m]}\bar{p}_{[n]}
 \cr
&&
C_{1\bar{1}}C_{(k)}p_{[m]}\bar{p}_{[n]}
=
\left(
(N+m+n-k-1)C_{1\bar{1}}C_{(k-1)}^{<1,\bar{1}>}
+C_{1\bar{1}}C_{(k)}^{<1,\bar{1}>}
\right)
p_{[m]}\bar{p}_{[n]} \cr
&&
C_{1\bar{1}}C_{(n)}p_{[m]}\bar{p}_{[n]}
=
(N+m+n-n-1)C_{1\bar{1}}C_{(n-1)}^{<1,\bar{1}>}p_{[m]}\bar{p}_{[n]}
\end{eqnarray}
Using these equations, we can show 
\begin{eqnarray}
C_{1\bar{1}}P_{[m]\bar{[n]}}
&=&
\left(
C_{1\bar{1}}+\sum_{k=1}^{n}(-1)^{k}\prod_{l=1}^{k}
\frac{C_{1\bar{1}}C_{(k)}}{(N+m+n-l-1)}
\right)
p_{[m]}\bar{p}_{[n]} \cr
&=&
\left(
C_{1\bar{1}}+\sum_{k=1}^{n}(-1)^{k}
\left(
f^{(k)}+f^{(k+1)}
\right)
\right)
p_{[m]}\bar{p}_{[n]} \cr
&=&0
\end{eqnarray}
where 
\begin{eqnarray}
f^{(k)}\equiv 
\prod_{l=1}^{k-1}
\frac{C_{1\bar{1}}C_{(k-1)}^{<1,\bar{1}>}
}{(N+m+n-l-1)}\hspace{0.2cm}(k=2,\cdots,n) 
\quad f^{(1)}\equiv C_{1\bar{1}} \quad f^{(n+1)}=0
\end{eqnarray}

In the same way, we can give a proof $C_{i\bar{j}}P_{R\bar{S}}=0$ for 
other cases. 
For antisymmetric representation, 
(\ref{sigmaonsymmetricrep}) gets replaced with 
\begin{equation}
\sigma p_{[1^{m}]}=-p_{[1^{m}]} 
\end{equation}
for any transposition $\sigma$. 


\subsection{Orthogonal set of gauge theory operators for some examples}
\label{symmetricoperators}
In this section, symmetric operators are listed for 
some examples: $(m,n)=(1,1), (2,1), (3,1)$ and $(2,2)$ 
using projectors in section \ref{examplesprojector}.
As we discussed in sections 
\ref{sec:projectorinBrauer} and 
\ref{sec:orthogonalsetoperator}, 
symmetric projectors for $k\neq 0$ 
are labelled by three indices 
$\gamma$, $A$ and $i$. The index $i$ runs over the multiplicity 
$M_{A}^{\gamma}$. 
For some examples we will consider here, 
the multiplicity takes $1$ 
for any $\gamma$ and $A$. 
Therefore we omit the index $i$ in this section. 
In the case of $k=0$, symmetric projectors 
are central element of the Brauer algebra and are 
labelled by an index $\gamma$. 
We express $P^{(k=0, \gamma_{+}=R, \gamma_{-}=S)}\equiv P_{R\bar{S}}$. 


\subsubsection{$m=1$, $n=1$}
\begin{eqnarray}
&&
tr_{(1,1)}(P_{[1]\bar{[1]}}\Phi)
=tr(\Phi)tr(\Phi^{\dagger}) 
-\frac{1}{N}tr(\Phi\Phi^{\dagger}) 
\nnm \\
&&
tr_{(1,1)}(P_{[1]\bar{[1]}}^{(k=1)}\Phi)
=\frac{1}{N}tr(\Phi\Phi^{\dagger}) 
\end{eqnarray}
To be precise, we should write 
$P_{[1]\bar{[1]}}^{(k=1,\gamma_{+}=\emptyset,\gamma_{-}=\emptyset)}$ 
instead of $P_{[1]\bar{[1]}}^{(k=1)}$. 
But we use the latter expression because there is only one choice 
for $k=1$ in this case, and this does not cause any confusion. 
We also use this notation for following examples. 

\subsubsection{$m=2$, $n=1$}
\begin{eqnarray}
&&
tr_{2,1}(P_{[2]\bar{[1]}}\Phi)
=\frac{1}{2}(tr\Phi)^{2}tr\Phi^{\dagger}
+\frac{1}{2}tr\Phi^{2}tr\Phi^{\dagger}
-tr_{2,1}(P_{[2]\bar{[1]}}^{(k=1)}\Phi)
\nnm \\
&&
tr_{2,1}(P_{[1^{2}]\bar{[1]}}\Phi)
=\frac{1}{2}(tr\Phi)^{2}tr\Phi^{\dagger}
-\frac{1}{2}tr\Phi^{2}tr\Phi^{\dagger}
-tr_{2,1}(P_{[1^{2}]\bar{[1]}}^{(k=1)}\Phi)
\nnm \\
&&
tr_{2,1}(P_{[2]\bar{[1]}}^{(k=1)}\Phi)
=\frac{1}{N+1}\left(tr(\Phi)tr(\Phi\Phi^{\dagger})
+tr(\Phi^{2}\Phi^{\dagger})\right) \nnm \\
&&
tr_{2,1}(P_{[1^{2}]\bar{[1]}}^{(k=1)}\Phi)
=\frac{1}{N-1}\left(tr(\Phi)tr(\Phi\Phi^{\dagger})
-tr(\Phi^{2}\Phi^{\dagger})\right)  
\end{eqnarray}

\subsubsection{$m=3$, $n=1$}
\begin{eqnarray}
&& 
 \hspace{-1.2cm}
tr_{3,1}(P_{[3]\bar{[1]}}\Phi)=
\frac{1}{6}(tr\Phi)^{3}tr\Phi^{\dagger}
+\frac{1}{2}tr\Phi^{2}tr\Phi tr\Phi^{\dagger}
+\frac{1}{3}tr\Phi^{3}tr\Phi^{\dagger}
-tr_{3,1}(P_{[3]\bar{[1]}}^{[2][\emptyset] (k=1)}\Phi) \nnm \\
&& 
 \hspace{-1.2cm}
tr_{3,1}(P_{[1^{3}]\bar{[1]}}\Phi)=
\frac{1}{6}(tr\Phi)^{3}tr\Phi^{\dagger}
-\frac{1}{2}tr\Phi^{2}tr\Phi tr\Phi^{\dagger}
+\frac{1}{3}tr\Phi^{3}tr\Phi^{\dagger}\nnm 
-tr_{3,1}(P_{[1^{3}]\bar{[1]}}^{[1^{2}][\emptyset] (k=1)}\Phi)
\nnm  \\
&& 
 \hspace{-1.2cm}
tr_{3,1}(P_{[2,1]\bar{[1]}}\Phi)=
\frac{2}{3}(tr\Phi)^{3}tr\Phi^{\dagger}
-\frac{2}{3}tr\Phi^{3}tr\Phi^{\dagger}
-tr_{3,1}(P_{[2,1]\bar{[1]}}^{[2][\emptyset] (k=1)}\Phi)
-tr_{3,1}(P_{[2,1]\bar{[1]}}^{[1^{2}][\emptyset] (k=1)}\Phi)
\nnm \\
&& 
\hspace{-1.2cm}
tr_{3,1}(P_{[3]\bar{[1]}}^{(k=1,\gamma_{+}=[2],\gamma_{-}=[\emptyset])}\Phi)
=
\frac{1}{N+2}\left(
\frac{1}{2}tr\Phi\Phi^{\dagger}(tr\Phi)^{2}
+\frac{1}{2}tr\Phi\Phi^{\dagger}tr\Phi^{2}
+tr\Phi^{2}\Phi^{\dagger}tr\Phi
+\frac{1}{3}tr\Phi^{3}\Phi^{\dagger}
\right) \nnm \\
&& \hspace{-1.2cm}
tr_{3,1}(P_{[1^{3}]\bar{[1]}}
^{(k=1,\gamma_{+}=[1^{2}],\gamma_{-}=[\emptyset])}\Phi)
=
\frac{1}{N-2}\left(
\frac{1}{2}tr\Phi\Phi^{\dagger}(tr\Phi)^{2}
-\frac{1}{2}tr\Phi\Phi^{\dagger}tr\Phi^{2}
-tr\Phi^{2}\Phi^{\dagger}tr\Phi
+\frac{1}{3}tr\Phi^{3}\Phi^{\dagger}
\right) \nnm \\
&& \hspace{-1.2cm}
tr_{3,1}(P_{[2,1]\bar{[1]}}^{(k=1,\gamma_{+}=[2],\gamma_{-}=[\emptyset])}\Phi)
=
\frac{1}{N-1}\left(
tr\Phi\Phi^{\dagger}(tr\Phi)^{2}
+tr\Phi\Phi^{\dagger}tr\Phi^{2}
-tr\Phi^{3}\Phi^{\dagger}
-tr\Phi^{2}\Phi^{\dagger}tr\Phi
\right) \nnm \\
&& \hspace{-1.2cm}
tr_{3,1}(P_{[2,1]\bar{[1]}}^{(k=1,\gamma_{+}=[1^{2}],\gamma_{-}=[\emptyset])}\Phi)
=
\frac{1}{N+1}\left(
tr\Phi\Phi^{\dagger}(tr\Phi)^{2}
-tr\Phi\Phi^{\dagger}tr\Phi^{2}
-tr\Phi^{3}\Phi^{\dagger}
+tr\Phi^{2}\Phi^{\dagger}tr\Phi
\right)\cr 
&&{\hspace{3in}}  
\end{eqnarray}
\subsubsection{$m=2$, $n=2$}
\begin{eqnarray}
&&
\hspace{-1.2cm}
tr_{2,2}(P_{[2]\bar{[2]}}\Phi)
=
\frac{1}{4}\left(
(tr\Phi)^{2}(tr\Phi^{\dagger})^{2}
+tr\Phi^{2}(tr\Phi^{\dagger})^{2}
+(tr\Phi)^{2}tr(\Phi^{\dagger})^{2}
+tr\Phi^{2}tr(\Phi^{\dagger})^{2}
\right) \nnm \\ 
&& \hspace{3cm}
-tr_{2,2}(P_{[2]\bar{[2]}}^{(k=1)}\Phi)
-tr_{2,2}(P_{[2]\bar{[2]}}^{(k=2)}\Phi)
\nnm \\
&&\hspace{-1.2cm}
tr_{2,2}(P_{[2]\bar{[1^{2}]}}\Phi)
=
\frac{1}{4}\left(
(tr\Phi)^{2}(tr\Phi^{\dagger})^{2}
+tr\Phi^{2}(tr\Phi^{\dagger})^{2}
-(tr\Phi)^{2}tr(\Phi^{\dagger})^{2}
-tr\Phi^{2}tr(\Phi^{\dagger})^{2}
\right) \cr
&& \hspace{3cm}
-tr_{2,2}(P_{[2]\bar{[1^{2}]}}^{(k=1)}\Phi)
  \nnm \\
&&
\hspace{-1.2cm}
tr_{2,2}(P_{[1^{2}]\bar{[2]}}\Phi)
=
\frac{1}{4}\left(
(tr\Phi)^{2}(tr\Phi^{\dagger})^{2}
-tr\Phi^{2}(tr\Phi^{\dagger})^{2}
+(tr\Phi)^{2}tr(\Phi^{\dagger})^{2}
-tr\Phi^{2}tr(\Phi^{\dagger})^{2}
\right) \cr
&& \hspace{3cm}
-tr_{2,2}(P_{[1^{2}]\bar{[2]}}^{(k=1)}\Phi)
  \nnm \\
&&
\hspace{-1.2cm}
tr_{2,2}(P_{[1^{2}]\bar{[1^{2}]}}\Phi)
=
\frac{1}{4}\left(
(tr\Phi)^{2}(tr\Phi^{\dagger})^{2}
-tr\Phi^{2}(tr\Phi^{\dagger})^{2}
-(tr\Phi)^{2}tr(\Phi^{\dagger})^{2}
+tr\Phi^{2}tr(\Phi^{\dagger})^{2}
\right) \cr
&& \hspace{3cm}
-tr_{2,2}(P_{[1^{2}]\bar{[1^{2}]}}^{(k=1)}\Phi)
-tr_{2,2}(P_{[1^{2}]\bar{[1^{2}]}}^{(k=2)}\Phi)
\nnm \\
&&
\hspace{-1.2cm}
tr_{2,2}(P_{[2]\bar{[2]}}^{(k=1)}\Phi)
=
\frac{1}{N+2}\left(
tr\Phi\Phi^{\dagger}tr\Phi tr\Phi^{\dagger}
+tr\Phi^{2}\Phi^{\dagger}tr\Phi^{\dagger}
+tr(\Phi(\Phi^{\dagger})^{2})tr\Phi
+tr\Phi^{2}(\Phi^{\dagger})^{2}
\right) \cr
&& \hspace{3cm}
-\frac{2}{N(N+2)}
\left(
(tr\Phi\Phi^{\dagger})^{2}+tr(\Phi\Phi^{\dagger}\Phi\Phi^{\dagger})
\right)
\nnm \\
&&
\hspace{-1.2cm}
tr_{2,2}(P_{[2]\bar{[1^{2}]}}^{(k=1)}\Phi)
=
\frac{1}{N}\left(
tr\Phi\Phi^{\dagger}tr\Phi tr\Phi^{\dagger}
+tr\Phi^{2}\Phi^{\dagger}tr\Phi^{\dagger}
-tr(\Phi(\Phi^{\dagger})^{2})tr\Phi
-tr\Phi^{2}(\Phi^{\dagger})^{2}
\right) \nnm \\
&&
\hspace{-1.2cm}
tr_{2,2}(P_{[1^{2}]\bar{[2]}}^{(k=1)}\Phi)
=
\frac{1}{N}\left(
tr\Phi\Phi^{\dagger}tr\Phi tr\Phi^{\dagger}
-tr\Phi^{2}\Phi^{\dagger}tr\Phi^{\dagger}
+tr(\Phi(\Phi^{\dagger})^{2})tr\Phi
-tr\Phi^{2}(\Phi^{\dagger})^{2}
\right) \nnm \\
&&
\hspace{-1.2cm}
tr_{2,2}(P_{[1^{2}]\bar{[1^{2}]}}^{(k=1)}\Phi)
=
\frac{1}{N-2}\left(
tr\Phi\Phi^{\dagger}tr\Phi tr\Phi^{\dagger}
-tr\Phi^{2}\Phi^{\dagger}tr\Phi^{\dagger}
-tr(\Phi(\Phi^{\dagger})^{2})tr\Phi
+tr\Phi^{2}(\Phi^{\dagger})^{2}
\right) \cr
&& 
\hspace{3cm}
-\frac{2}{N(N-2)}
\left(
(tr\Phi\Phi^{\dagger})^{2}-tr(\Phi\Phi^{\dagger}\Phi\Phi^{\dagger})
\right)
 \nnm \\
&&
\hspace{-1.2cm}
tr_{2,2}(P_{[2]\bar{[2]}}^{(k=2)}\Phi)
=
\frac{1}{N(N+1)}\left(
(tr\Phi\Phi^{\dagger})^{2}+tr(\Phi\Phi^{\dagger}\Phi\Phi^{\dagger})
\right)
\nnm \\
&&
\hspace{-1.2cm}
tr_{2,2}(P_{[1^{2}]\bar{[1^{2}]}}^{(k=2)}\Phi)
=
\frac{1}{N(N-1)}\left(
(tr\Phi\Phi^{\dagger})^{2}-tr(\Phi\Phi^{\dagger}\Phi\Phi^{\dagger})
\right)
\end{eqnarray}

\end{appendix}

\end{document}